\documentclass[]{aa} 
\makeatletter
\renewcommand*\aa@pageof{, page \thepage{} of \pageref*{LastPage}}
\makeatother
\usepackage{graphicx,comment}
\usepackage{txfonts}
\usepackage{newtxtext,newtxmath}
\usepackage{amsmath}	

\usepackage{anyfontsize}
\usepackage{xspace}
\usepackage{amssymb}
\usepackage{amsmath} 
\usepackage{graphicx}
\usepackage{txfonts}
\usepackage{lipsum}
\usepackage{longtable}

\newcommand{\thisgrb}{GRB 230307A\xspace}
\newcommand{\astr}{{\em AstroSat}\xspace}
\newcommand{\frm}{{\em Fermi}\xspace}
\newcommand{\kw}{{\em Konus}-\textit{Wind}\xspace}
\newcommand{\kwT}{T$_{\rm kw,0}$}

\newcommand{\tfr}{{T$_{\rm 0}$}\xspace}
\newcommand{\en}{{\rm keV}\xspace}

\newcommand{\tnty}{{$T_{\rm 90}$}\xspace}

\begin{document}

\title{Time-resolved spectro-polarimetric analysis of extremely bright GRB 230307A: Possible Evidence of evolution from photospheric to synchrotron dominated emission}
\titlerunning{Spectro-polarimetric analysis of GRB 230307A}
\authorrunning{Gupta et al.}

\author{Soumya Gupta\inst{1, 2}\fnmsep\thanks{Corresponding author: Soumya Gupta}
\and Rahul Gupta\inst{3,4}\fnmsep\thanks{Corresponding author:  Rahul Gupta}
\and Tanmoy Chattopadhayay\inst{5}
\and Sunder Sahayanathan\inst{1,2}
\and D. Frederiks \inst{6}
\and D. Svinkin \inst{6}
\and Dipankar Bhattacharya\inst{8}
\and Judith Racusin \inst{3}
\and Santosh Vadawale\inst{7}
\and Varun Bhalerao\inst{9}
\and A. Lysenko \inst{6}
\and A. Ridnaia \inst{6}
\and A. Tsvetkova \inst{6, 10}
\and M. Ulanov \inst{6}
}

\institute{Bhabha Atomic Research Center, Mumbai, Maharashtra-400094, India\\ \email{soumya.gupta1512@gmail.com, soumyag@barc.gov.in}
\and Homi Bhabha National Institute, Mumbai, Maharashtra-400094, India   
\and Astrophysics Science Division, NASA Goddard Space Flight Center, Mail Code 661, Greenbelt, MD 20771, USA \\ \email{rahulbhu.c157@gmail.com} 
\and NASA Postdoctoral Program Fellow 
\and Kavli Institute of Particle Astrophysics and Cosmology, Stanford University, 452 Lomita Mall, Stanford, CA 94305, USA 
\and Ioffe Institute, 26 Politekhnicheskaya, St. Petersburg, 194021, Russia
\and Physical Research Laboratory Thaltej, Ahmedabad, Gujarat 380009, India
\and Department of Physics, Ashoka University, Sonipat, Haryana-131029, India  
\and Indian Institute of Technology Bombay, Powai, Mumbai, Maharashtra 400076, India 
\and Department of Physics, University of Cagliari, SP Monserrato-Sestu,
km 0.7, 09042 Monserrato, Italy }

\date{Received 2025; accepted}

\abstract
{The radiation mechanisms powering Gamma-ray bursts (GRBs) and their physical processes remain one of the unresolved questions in high-energy astrophysics. Spectro-polarimetric observations of exceptionally bright GRBs provide a powerful diagnostic tool to address these challenges.} 
{GRB 230307A, the second-brightest long-duration GRB ever detected, exhibits a rare association with a Kilonova, offering a unique and rare probe into the emission processes of GRBs originating from compact object mergers.}
{We present a comprehensive time-averaged and time-resolved spectro-polarimetric analysis of GRB 230307A using joint observations from the \astr Cadmium Zinc Telluride Imager (CZTI), the \frm Gamma-ray Burst Monitor (GBM), and \kw.} 
{Spectral analysis reveals a temporal evolution in the low-energy photon index, $\alpha$, transitioning from a hard to a softer state over the burst duration. Time-averaged polarimetric measurements yield a low polarization fraction ($<$ 12.75 \%), whereas time-resolved polarization analysis unveils a marked increase in polarization fractions ($>$ 49.0 \%) in the later stages of the emission episode.}
{This spectro-polarimetric evolution suggests a transition in the dominant radiative mechanism: the initial phase, possibly characterized by thermal-dominated photospheric emission (unpolarized or weakly polarized), gives way to a regime dominated by non-thermal synchrotron emission (highly polarized). This transition provides possible evidence for the evolving influence of magnetic fields in shaping the GRB emission process and jet dynamics.}


\maketitle

\section{Introduction}

The understanding of the composition of the relativistic jets, as well as the mechanism responsible for the prompt gamma-ray emission from the gamma-ray bursts (GRB), is still some of the most profound and unresolved questions \citep{peer_08, Beloborodov12, 2015AdAst2015E..22P, zhang14, 2022Galax..10...38B}. Several competing theoretical models have been proposed to elucidate the processes responsible for prompt emission. One of the prevailing models explaining GRB prompt emission involves the internal shocks within the jet, where relativistic shells of plasma collide, releasing energy in the form of gamma rays \citep{rees94, Tavani1996, Bonjak2009,2024MNRAS.528L..45R}. This internal shock model can account for the observed temporal variability and the spectral properties of the prompt emission \citep{1997ApJ...490...92K, 1997MNRAS.287..110S}. Alternatively, the photospheric emission model suggests the photospheric emission boosted to gamma-rays originates as thermal radiation at the photosphere, providing insights into the jet's baryonic composition and opacity \citep{2005ApJ...628..847R, 2006A&A...457..763G, 2011MNRAS.415.1663T, Gupta_2024}. 

Traditionally, the spectral characteristics of prompt emissions are utilized to examine the radiation mechanisms \citep{2019A&A...625A..60R, 2021MNRAS.505.4086G, 2023MNRAS.519.3201C, 2024A&A...683A..55C}. However, the existence of degeneracy among various spectral models, all yielding equally viable statistics, underscores the need for an additional observational constraint, for example, polarization \citep{2018JApA...39...75I,jirong_hybrid_pol}. The measurement of polarization is particularly crucial because distinct radiation models predict different levels of polarization fractions based on the orientation of the emitting jet \citep{2021Galax...9...82G}. Additionally, polarimetry observations also offer unique insights into the relativistic outflow's composition (baryonic jets, dominated by the kinetic energy of particles, or Poynting-flux-dominated jets, where magnetic fields primarily carry the energy), and angular geometry of the relativistic jets \citep{2025ApJS..276....9L,mao13,jirong_hybrid_pol}. As a result, combining spectral and polarization analyses proves highly valuable for comprehending the radiation mechanisms underlying prompt emissions \citep{Sharma20, 2022MNRAS.511.1694G}. The measurement of time-integrated polarization of prompt emission is conducted, primarily owing to the substantial photon requirements of X-ray polarimetry \citep{chattopadhyay21_review, Kole20polar_catalog}. However, time-integrated polarization measurement of GRBs may result in low polarization due to the cancellation of polarization vectors \citep{chatt_22_sample}. Therefore, detailed time-resolved polarization measurements are needed to constrain the intrinsic radiation physics of GRBs\citep{Burgess_etal_2019, 2024ApJ...972..166G, yonetoku11, gotz09, Sharma_etal_2019}.

In this paper, we present a comprehensive spectral and polarimetric analysis of GRB 230307A using data from the Cadmium Zinc Telluride Imager (CZTI) aboard \textit{AstroSat} \citep{2014SPIE.9144E..1SS}, Gamma-ray Burst Monitor (GBM) aboard \emph{Fermi} \citep{2009ApJ...702..791M}, and \kw (KW) \citep{1995SSRv...71..265A} instrument. Notably, this event represents the brightest burst for which spectro-polarimetric analysis of the prompt emission of a GRB is performed. In Section \ref{obs}, we describe the high-energy observations of GRB 230307A and the data reduction methodologies employed. Sections \ref{result} and \ref{diss} present the principal results of our analysis and discuss the implications. Finally, Section \ref{summary} provides a succinct summary and conclusions of the present work.

\section{Prompt emission observations and data analysis} \label{obs}

\subsection{Data collection}

\subsubsection{Fermi/GBM}
\thisgrb was first reported by the GBM onboard \frm satellite at 15:44:06 UT on March 23, 2023 (hereafter \tfr; \citealt{gcn_fermi}). The burst exhibited a fluence of 2.951 $\pm$ 0.004 $\rm 10^{-3} erg/cm^2$ in the 10-1000 keV energy range \citep{gcn_fermi_fluence}, making it the second brightest burst ever detected. Temporal analysis of the GBM data revealed a single-peaked structure (see Fig. \ref{param_all}) with the \tnty duration of 35 s in the 10-1000 \en energy range. For this study, we utilized the Time-Tagged Event (TTE) data from the brightest Sodium Iodide (NaI 10: 52$^\circ$, 8-900 keV) and bismuth germanate detector (BGO 1: 60$^\circ$, 0.3-40 MeV) detectors of the instrument GBM. Due to the extreme brightness of GRB 230307A, the GBM detectors experienced significant pulse pile-up during specific intervals. Consequently, data from 2.5–7.5 s in the NaI detector and 2.5–11.0 s in the BGO detector were excluded from the analysis to ensure the reliability of the spectral and temporal measurements.

\subsubsection{AstroSat-CZTI}
GRB 230307A was identified as the brightest burst detected by the CZTI onboard \astr. The light curve of the event exhibited a single-peaked structure, with the peak occurring at 15:44:10.0 UT and a duration (\tnty) of 33 seconds in the 20-200 keV energy range. Preliminary analysis revealed 5484 polarization events associated with the burst \citep{gcn_czti}. The incident direction of GRB 230307A, as observed by CZTI, was determined to be at polar and azimuthal angles of ($\theta$ = 150.46$^\circ$) and ($\phi$ = 185$^\circ$), respectively. 

\astr-CZTI has been established as a sensitive on-axis GRB polarimeter, with its performance validated during ground calibration in the 100-350 keV energy range \citep{chatt_2014, vada_2015}. Recent advancements have extended its polarimetric measurement capabilities for off-axis and in the 100-600 keV energy range \citep{chatt_22_sample, 2024ApJ...972..166G}. We utilized this enhanced CZTI's capabilities in probing the prompt emission polarization properties of GRB 230307A (details in \S\ref{polan}).

\subsubsection{Konus-Wind}
KW also reported the detection of the \thisgrb at 15:44:05.615 UTC on 7 March 2023 (hereafter referred to as \kwT). The KW trigger time corresponds to the Earth-crossing time of 15:44:06.667 UTC \citep{gcn_konus}. The peak count rate measured by KW for this burst was $\sim 1\times 10^5$ counts per second, which is an order of magnitude lower than the count rate observed for the exceptionally bright GRB 221009A, also known as the ``BOAT" (Brightest Of All Time) event \citep{Frederiks_2023ApJ_949L_7}. At this count rate, the KW light curve remains unaffected by saturation or pulse-pileup effects, allowing for the application of standard KW dead-time correction techniques \citep{Mazets_1999AstL_25_635, Frederiks_2023ApJ_949L_7} to the data. Thus, both \frm/GBM and KW data were utilized to carry out the time-resolved spectral analysis (detailed analysis described in Sect. \ref{spec_temp}), to draw a comprehensive inference regarding the emission mechanism of \thisgrb.

In addition to these instruments, the GRB was detected by several other space-based instruments operating in the $\gamma$-ray and hard X-ray regimes. These include the Gravitational Wave High-Energy Electromagnetic Counterpart All-sky Monitor (GECAM) \citep{gcn_gecam}, GRBAlpha \citep{gcn_alpha}, and the Gamma-Ray Imaging Detector (GRID) and Mini-CALorimeter (MCAL) aboard the AGILE satellite \citep{gcn_agile}. 

\subsection{Temporal and spectral analysis}\label{spec_temp}

\begin{figure}
    \centering
    \includegraphics[width = 0.5\textwidth]{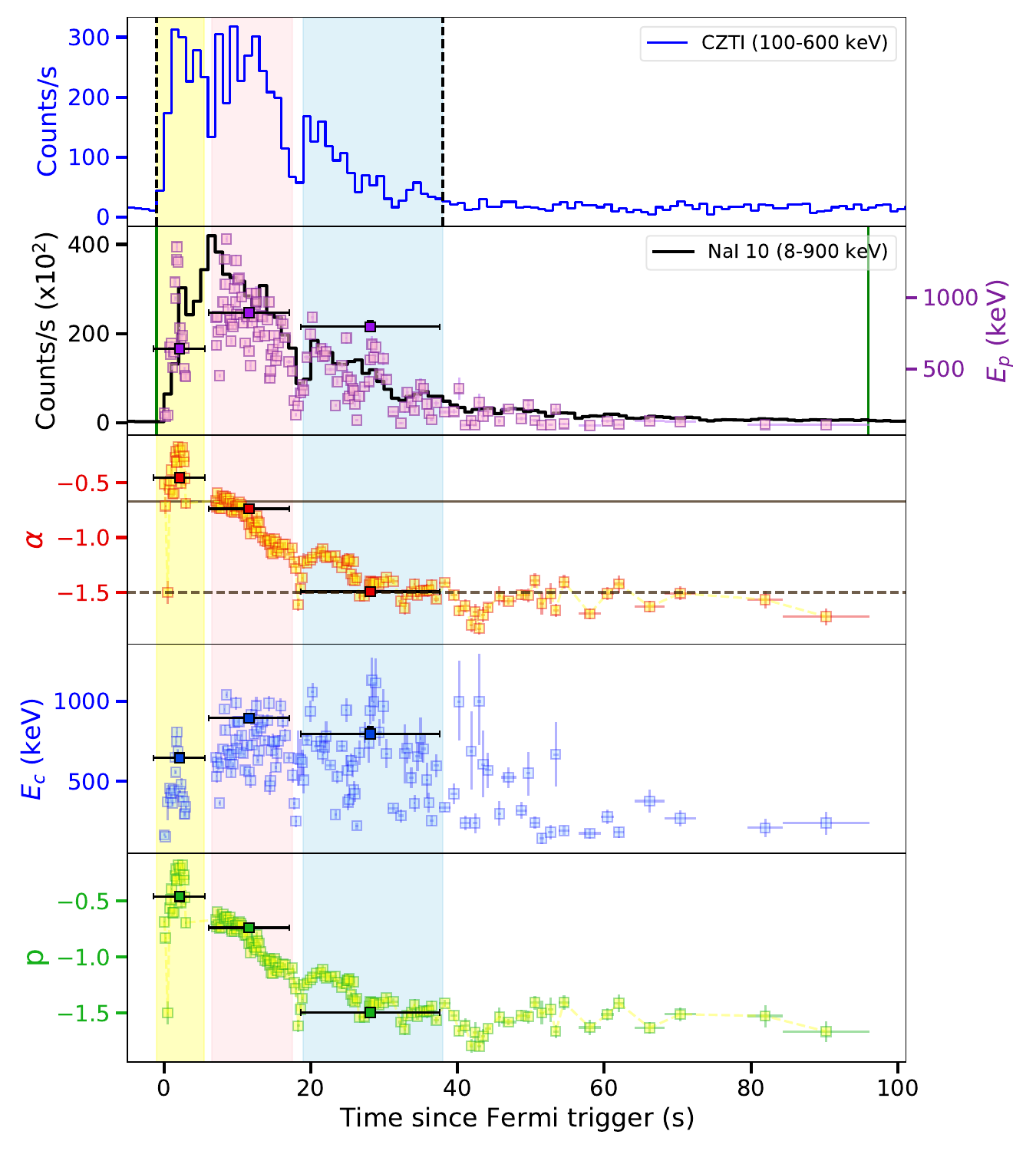}
    \caption{The first panel shows the Compton light curve of CZTI data in the 100-600 keV energy range, 
    with the black dashed line indicating the burst interval determined using the Bayesian block technique. 
    The second panel illustrates the light curve from the NaI 10 detector (black), and the green solid lines 
    depict the burst interval of \frm GBM using the Bayesian block technique. 
    The second through fifth panels illustrate the temporal evolution of spectral 
    parameters throughout the burst (Band function parameters: peak energy $E_p$, low-energy spectral index $\alpha$, and CPL parameters: cut-off energy $E_c$, power-law index $p$). Spectral parameters derived from Konus-Wind data are shown as dark markers 
    with black error bars, while the finer time-bin spectra modeled using \frm data are displayed as lighter 
    background points. In the third panel, the solid and dashed brown lines indicate the synchrotron limits for 
    slow-cooling (-2/3) and fast-cooling (-3/2) scenarios, respectively. Shaded regions highlight the time-resolved intervals used for the polarization analysis.}
    \label{param_all}
\end{figure}

The \frm GBM light curve was extracted using the RMFIT software (version 4.3.2). Figure \ref{param_all} presents the background-subtracted light curves from the NaI 10 in 8-900 keV (panel 2, black curve). The time intervals affected by pulse pile-up (NaI: 2.5–7.5 s; BGO: 2.5–11.0 s) are excluded from subsequent spectral analysis. The top panel displays the \astr CZTI light curve in the energy ranges of 100–600 keV (blue). 

Spectral analysis of GRB 230307A was carried out using data from the KW and \frm/GBM instruments. Three time-resolved KW spectra were extracted and analyzed, with temporal bins determined using the Bayesian blocks technique applied to the CZTI light curve (details provided in Sect. \ref{polan}). The spectra in the first two regimes include the time intervals where the GBM spectra cannot be studied due to the pulse pile-up. The KW spectral analysis was performed using \texttt{XSPEC} (version 12.11.1) with the $\chi^2$ statistic. To ensure the validity of the $\chi^2$ statistic, energy channels were grouped to achieve a minimum of 20 counts per channel. The KW instrument provided 64-channel energy spectra via two pulse-height analyzers: PHA1 (63 channels, 29–1660 keV) and PHA2 (60 channels, 0.5–20.3 MeV). 

The extreme brightness of GRB 230307A enabled high-time-resolution spectral analysis using \frm/GBM data. The Bayesian block technique was applied to the GBM light curve over the total emission interval (\tfr to \tfr + 96.1 s), yielding 150-time bins with high statistical significance (see Table \ref{table_spec}). Spectra for these 150 intervals were generated using the \texttt{Make spectra} tool within the \texttt{gtburst} software from the Fermi Science Tools. Background estimation was performed by selecting two-time intervals, one preceding and one following the main GRB emission. Time-averaged spectra were modeled using the Multi-Mission Maximum Likelihood framework (3ML) to investigate potential emission mechanisms. The 33–40 keV energy range was excluded from the analysis due to the presence of the iodine K-edge at 33.17 keV. The model selection was guided by the Bayesian information criterion (BIC, \citealt{1978schw}), with preference given to models yielding the lowest BIC values.

\begin{table*}
\begin{center}
\caption{The polarisation and spectral results of GRB 230307A using \astr-CZTI (100–600 keV) and \kw (20 keV–20 MeV) respectively.}
 \label{pol_tab}
 \begin{tabular}{|c|c|c|c|c|c|c|c|c|c|c|}
   \hline 
   $ T_{i}-T_{f}$ & {Compton} & {Modulation}  & {CZTI PA/(sky PA)} &{PF$^{*}$}&BF&{$\alpha$}&{$E_p$}&{$\beta$}\\
    (sec)& {Counts} & {Amplitude}  &  $^{\circ}$& (\%)& &keV & &\\

  \hline 
-1.0-38.0& 5122& $0.06\pm0.05$ & - & $<12.75$ & 0.97& $-0.89_{-0.01}^{+0.01}$& $1052_{-8.12}^{+8.45}$& $<-5.9$\\ &&&&&&&&\\
\hline
-1.0-6.0 & 1542& $0.07\pm0.06$ & - & $<24.59$ & 0.73&$-0.45_{-0.01}^{+0.01}$ & $989.0_{-12.0}^{+12.0}$ & $-4.98_{-0.64}^{+0.37}$ \\

6.5-17.5 & 2434& $0.04\pm0.04$ &-  & $<19.40$ & 0.69&$-0.74_{-0.01}^{+0.01}$ & $1127.0_{-13.0}^{+13.0}$ & $<-5.2$ \\

19.0-38.0& 1067& $0.19\pm0.08$ & $85.76\pm25.00$  (8.17)&$>49.0$ & 4.62&$-1.49_{-0.01}^{+0.01}$ & $403.0_{-11.0}^{+12.0}$ & $<-4.4$ \\

\hline
\end{tabular}
\end{center}
$^*$ The upper and lower limits are computed at 1$\sigma$ level for two and one parameter of interest, respectively.
\end{table*}

\subsection{Polarimetric analysis}\label{polan}

The spectral analysis alone often results in ambiguity when selecting the best-fit model, highlighting the need for additional constraining observables such as polarization to resolve such degeneracies. Polarization measurement depends on the cross-section of X-ray interaction with matter, which modulates the intensity of photons or electrons as a function of polarization direction \citep{mcconnell16}. The pixelated CZTI detectors onboard \astr are uniquely capable of detecting ionizing events in neighboring pixels simultaneously, enabling the reconstruction of the azimuthal distribution of Compton-scattered events. This capability allows CZTI to function as a Compton polarimeter, providing polarization measurements of incoming hard X-ray radiation (above 100 keV). 

In CZTI, Compton scattering events are identified by detecting adjacent two-pixel events within a 20 $\mu$s time window, with energy ratios between 1 and 6 to reduce noise. Data from both GRB and background intervals are analyzed, and the azimuthal scattering angle is calculated. Corrections for the asymmetric inherent pixel geometry of CZTI are applied using Geant4 simulations of the AstroSat mass model, validated by \cite{mate21, chattopadhyay21_grb}. The corrected azimuthal distributions are fitted with a sinusoidal function to derive the modulation amplitude and polarization angle. The polarization fraction is then calculated by normalizing the modulation amplitude to the simulated response for 100\% polarized radiation. A Bayes factor (BF) is employed to confirm polarization detection (BF $> 3.2$), where the Polarization fraction (PF) and Polarization angle (PA) are quoted for the detections. However, during the time bins where the fewer Compton counts registered, the PF can not be constrained due to the poor statistics, and thus, a lower limit on the PF is quoted. The lower limits are calculated at two parameters of interest, from the contour plot of PF and detector PA (fourth panel of the right-bottom image in Figure \ref{pol-res}). Whereas for the intervals with non-detections (BF $< 3.2$), the upper limit is reported at one parameter of interest. The upper limits are computed from the probability (n sigma) of a known 100\% unpolarized radiation exceeding a certain false polarization level that satisfies the detection criteria BF $>3.2$ \citep{chatt_22_sample}.
\begin{figure}
    \centering
    \includegraphics[width = 0.5\textwidth]{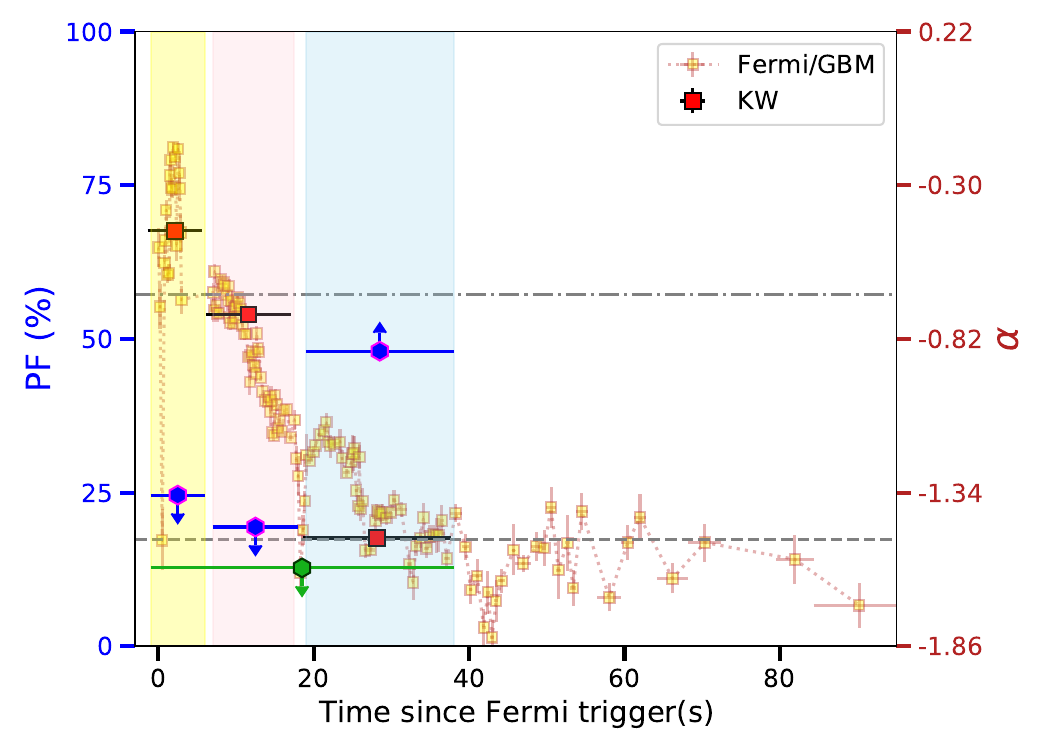}
    \caption{Time-averaged and time-resolved spectro-polarimetric analysis of GRB 230307A: The plot illustrates the evolution of the polarization fraction (PF) in 100-600 keV during the time-resolved (blue hexagon markers) and time-integrated (green hexagon marker) intervals of GRB 230307A. Three distinct time-resolved regions are highlighted with yellow, red, and blue shaded areas. The right vertical axis presents the corresponding variation in the low-energy spectral index ($\alpha$) derived from the Band function fits. The spectral indices obtained using time-resolved spectral analysis of \frm data are represented by orange squares, while those from \kw data (for each region) are denoted by red squares. The grey dashed and dot-dashed lines indicate the theoretical bounds for the fast-cooling and slow-cooling synchrotron emission regimes, respectively, providing a reference for evaluating the observed spectral behavior.}
    \label{polalp}
\end{figure}

For the polarimetric analysis of GRB 230307A, we also consider recent advancements such as extending the energy range to 600 keV by utilizing low-gain pixels, implementing a Compton noise algorithm for improved noise rejection, correcting for charge-sharing effects, and validating the AstroSat mass model \citep{chatt_22_sample}. These refinements have significantly enhanced the polarimetric sensitivity of \astr and mitigated systematic effects, ensuring robust polarization measurements. However, \astr-CZTI data for GRB 230307A were affected by saturation above the detector threshold due to the extreme brightness of the event, which introduced distortions in standardized polarization analysis. Approximately 10\% of Compton events were misidentified as triple-pixel events, while $\sim$ 10–20\% of single-pixel events were erroneously classified as Compton events. This is corrected by identifying and removing the pile-up regions in the CZTI. For that, we analyze the individual quadrant light curves (binned at 10 ms), where time intervals with event rates exceeding 30 counts per 10 ms were flagged as contaminated and excluded. Livetime corrections were performed for these events for each quadrant, reducing the total burst duration by a factor of 0.637. Polarization analysis was conducted for individual quadrants, yielding consistent results that were combined to determine the time-integrated and time-resolved polarization fraction (PF) and polarization angle (PA), as discussed in subsequent sections. The minimum detectable polarization (MDP) was calculated from the Geant4 simulation of the \astr mass model, to obtain the modulation amplitude for the 100\% polarised radiation for the GRB. Based on the number of Compton scattered photons, the MDP threshold computed for the total burst duration was 10\%, while 22\% for Region 1, 18\% for Region 2, and 30\% for Region 3. In this study, the uncertainties in the PA are reported at the 68\% confidence level for two parameters of interest, while the uncertainties in the PF are quoted at the 68\% confidence level for one parameter of interest unless otherwise specified. 

The Bayesian block technique \citep{Scargle2013} was applied to the Compton event light curve of the CZTI to determine the total duration of GRB 230307A for polarization analysis in the 100–600 keV energy range. The temporal evolution of polarization during the prompt emission phase of GRB 230307A may nullify polarization features when analyzed over the full burst duration, necessitating time-resolved polarization studies. Three distinct temporal intervals, identified using the Bayesian block technique, were selected for detailed time-resolved polarization analysis: Region 1 ($T_0$–1.0 s to $T_0$+6.0 s), Region 2 ($T_0$+6.5 s to $T_0$+17.5 s), and Region 3 ($T_0$+19.0 s to $T_0$+38.0 s). These intervals are highlighted as shaded regions in Figures \ref{param_all} and \ref{polalp}. 

\section{Results}\label{result}

\subsection{Time-averaged spectro-polarimetric properties}

The time-integrated spectrum of GRB 230307A ($T_0$-1.4 to $T_0$+37.6 s) was modelled using two empirical functions: the Band function \citep{band93} and a power law with exponential cut-off (CPL). Analysis of \kw spectrum within the 20–20000 keV energy range revealed that the Band function provided the best fit, yielding a low-energy spectral index $\alpha$ of $-0.89_{-0.01}^{+0.01}$, an upper limit on the high-energy spectral index $\beta$ of $-5.9$, and peak energy $E_{\text{p}}$ of $1052_{-8.12}^{+8.45}$ keV. The Bolometric flux was calculated as $\log F_{\text{bol}} = 3.922_{-0.001}^{+0.001}$ ergs/cm$^2$/s, confirming this burst to be highly energetic.

GRB 230307A, also the brightest burst detected by \astr/CZTI, recorded the highest Compton counts of 5122 during 39 seconds ($T_0$-1.0 to $T_0$+38.0 s), which is highlighted by the black-dashed lines in the top panel of Figure \ref{param_all}. Polarization analysis over the entire burst duration revealed a non-polarized emission with an upper limit on the polarization fraction of 12.75\% (green hexagon in Figure \ref{polalp}). Details of the upper limit calculation can be found in \citep{chatt_22_sample}. The top-left panel of Figure \ref{pol-res} provides the results of the time-averaged polarimetric analysis using CZTI data. The complete-burst analysis showed a low polarization signature, suggesting that a 
more detailed time-resolved analysis is necessary (see Table \ref{pol_tab}). Such analysis can reveal if the burst is intrinsically unpolarized (due to photospheric emission) or if observed low polarization is due to evolution in PA and PF.

\subsection{Evolution of polarization and spectral characteristics of GRB 230307A}

The time-resolved spectral analysis of prompt emission offers crucial insights into the underlying radiation processes and correlations between different spectral parameters. Time-resolved spectral modelling of \thisgrb was performed using data from \frm/GBM (light points in Figure \ref{param_all}) and \kw (dark points in shaded regions of Figure \ref{param_all}), using the Band function \citep{band93} and a power-law with an exponential cut-off (CPL). The GBM spectrum was analyzed across energy ranges of 8–900 keV (NaI) and 250–40000 keV (BGO). The Band function consistently provided the best fit ($\Delta$ BIC $<$ -10) in most of the time stamps, as indicated by lower Bayesian Information Criterion (BIC) values. The best-fit model parameters are quoted in the Table \ref{table_spec}.

The evolution of the spectral peak energy was found to follow a similar evolution trend as the flux over the burst duration. Additionally, the high-energy index $\beta$ was constant throughout the burst duration. On the other hand, the low-energy spectral index ($\alpha$) exhibited a distinct hard-to-soft evolution, indicative of a radiative process transition from thermal to non-thermal dominance. In Region 1 (yellow-shaded region of Figure \ref{param_all}), $\alpha$ was measured to be $-0.45\pm0.01$, significantly harder than the synchrotron ``line of death" ($\alpha \approx -2/3$; \citealt{Zhang_12_lod}). Regions 2 and 3 (red and blue shaded regions) showed $\alpha$ values of $-0.74\pm0.01$ and $-1.49\pm0.01$, respectively (see Table \ref{pol_tab}).

The extraordinary brightness of GRB 230307A provided a rare opportunity to investigate the evolution of its polarization properties across different phases of the burst. The initial emission during region 1 ($T_0$-1.0 s to $T_0$+6.0 s) with 1542 Compton counts exhibited no significant polarization, with a Bayes factor of 0.73 (BF $<$ 3; top-right image of Figure \ref{pol-res}). Similarly, the emission from region 2 ($T_0$-6.5 s to $T_0$+17.5 s) was unpolarized, showing 2434 Compton counts and a BF of 0.69 (bottom-left image of Figure \ref{pol-res}). The upper limits (at 1 $\sigma$ confidence level) on the polarization fraction for regions 1 and 2 were constrained to 24.59\% and 19.40\%, respectively. Interestingly, in the decaying phase of the light curve (region 3), the emission became distinctly polarized, with a lower limit on the PF of 49.0\% (blue hexagon scatterer in Figure \ref{polalp}) and a polarization angle of $85.76\pm25.0^\circ$ (bottom-right image of Figure \ref{pol-res}). This evolution, depicted in Figure \ref{polalp}, demonstrates a lower polarization fraction during the initial phases of the burst and a significantly higher polarization signature in the later phase, suggesting a potential change in the emission mechanism over time.

\section{Discussion}\label{diss}
The observed polarization fraction in GRBs is highly sensitive to the underlying radiation mechanisms. For instance, synchrotron radiation from a highly ordered magnetic field is expected to produce high polarization fractions, whereas a tangled or multi-zone magnetic field configuration would result in a significantly lower polarization ($\leq$ 20\%) \citep{2021Galax...9...82G}. Recent theoretical advancements suggest that the presence of a photospheric component, which is thermal in nature, can further complicate the polarization signature by diluting the non-thermal synchrotron emission \citep{lundman14}. 

The spectro-polarimetric analysis of GRB 230307A, based on data from the \astr/CZTI and \frm/GBM instruments, offers valuable insights into the prompt radiation mechanisms of this exceptionally energetic burst. The observed temporal evolution of polarization properties, combined with spectral characteristics, reveals a hint of transition in the radiation mechanisms from thermal to non-thermal dominance (see details in the following sub-sections). However, due to the statistical limitations of the current dataset, we present these interpretations as possible scenarios suggested by the measurements, rather than definitive conclusions.

The viewing geometry of a GRB jet also plays a pivotal role in determining the observed polarization properties. The $\Gamma \theta_{\rm j}$ condition, where $\Gamma$ is the bulk Lorentz factor and $\theta_{\rm j}$ is the jet opening angle, provides a robust framework for assessing whether the burst is observed on-axis ($\Gamma \theta_{\rm j}$ $>>$ 1) or off-axis ($\Gamma \theta_{\rm j}$ $<<$ 1), or within a narrow jet ($\Gamma \theta_{\rm j}$ $\sim$ 1). For GRB 230307A, we derive $\Gamma$ using the well-established Liang correlation, $\Gamma_{0}$ $\approx$ 182 $\times$ $E_{\gamma, \rm iso, 52}^{0.25 \pm 0.03}$ which relates the bulk Lorentz factor to the isotropic gamma-ray energy \citep{2010ApJ...725.2209L}. The jet opening angle $\theta_j$ of $3.4^\circ$ is calculated using afterglow analysis \citep{zhang_230307_thetaj} and $E_{\gamma, \rm iso, 52}$ is obtained from \cite{2024ApJ...969...26P}. Our analysis reveals that GRB 230307A has $\Gamma \theta_{\rm j}$ $>>$ 1, indicating an on-axis viewing perspective. This is consistent with the bright prompt emission and well-detected afterglow, which are characteristic of on-axis observations. The on-axis geometry implies that the observed polarization is less likely to be influenced by geometric effects such as jet edge contributions, allowing for a more direct interpretation of the intrinsic polarization properties. Below, we discuss the implications of our findings in the context of theoretical models and their relevance to the broader understanding of GRB physics.

\subsection{Initial phase: photospheric emission}

In the first temporal bin of GRB 230307A ($T_0$-1 – $T_0$+6 s), we observe a moderate polarization fraction (PF $<$ 24.59\%, 1 $\sigma$) alongside a hard low-energy spectral index ($\alpha$ $\sim$ -0.45), as shown in Figure \ref{polalp}. The low PF suggests that the emission is either intrinsically unpolarized or significantly depolarized, potentially due to a tangled magnetic field in the emitting region, where synchrotron emission is expected to dominate \citep{2021Galax...9...82G}. The hard $\alpha$ during the peak of GRB 230307A indicates a spectrum above the slow ($\alpha$ = -2/3) and fast-cooling synchrotron limit ($\alpha$ = -3/2), suggesting that such hard $\alpha$ could not be explained with a typical thin shell synchrotron emission model. However, several intricate theoretical models have been suggested that can generate a hard $\alpha$ (as seen during the peak emission phase of GRB 230307A) within the context of the synchrotron emission framework. These include synchrotron emission in a decaying magnetic field \citep{2006ApJ...653..454P, 2014NatPh..10..351U}, time-varying cooling of synchrotron electrons \citep{2018MNRAS.476.1785B, 2020NatAs...4..174B}, among others. Alternatively, this could reflect a photospheric component (the optical depth drops to unity, allowing photons to escape), such as proposed for GRB 090902B, which exhibited a hard $\alpha$ \citep{ryde2010, 2000ApJ...530..292M, peer_08, Gupta_2024}. In this scenario, the thermal radiation is expected to be unpolarized due to the random orientation of photons in the optically thick region \citep{lundman14}. The low-energy spectral index during this phase further supports a thermal origin, aligning with predictions of the photospheric model \citep{ryde2010}. \cite{2023ApJ...953L...8W} also analyse the prompt emission of GRB 230307A and noted a correlation between $\alpha$ with intensity. They further explain this correlation in terms of the evolution of the ratio of thermal to non-thermal components. The low PF in Region 1 aligns with this interpretation, as the photospheric contribution could suppress polarization, while the hard $\alpha$ reflects a thermal peak. Similar behavior was reported in a study by \cite{Sharma20} where it was indicated as a common early-phase depolarization in GRBs with mixed emission mechanisms. In a plausible scenario, the low-PF observed in the early phase, suggestive of photospheric emission, could align with the hard $\alpha$ values derived from spectral analysis of early emission phases \citep{2021ApJS..254...35L}. However, the observed $\alpha$–intensity correlation may also be a natural consequence of the commonly observed hard-to-soft spectral evolution in GRB pulses \citep{2011ApJ...740..104H, 1995ApJ...439..307F, 2006ApJS..166..298K}. During this evolution, the spectrum typically softens over time, with $\alpha$ decreasing as the intensity peaks and subsequently declines. This temporal behaviour, well-studied in many GRBs, could naturally account for the correlation without invoking a dominant photospheric component.

\subsection{Intermediate phase: sustained low polarization fraction with softening spectral index}

In the second temporal bin (6.5–17.5 s), the polarization fraction remains low (PF $<$ 19.40\%, 1 $\sigma$), while the spectral index softens to $\alpha$ $\sim$ -0.73, indicating an initiation of transition in the emission properties. The gradual softening $\alpha$ from region 1 to region 2, approaching below the slow-cooling synchrotron limit ($\alpha$ = -2/3), possibly suggests dilution of the thermal plasma further cooling it down. This weakens the photospheric emission and the spectrum may be dominated by synchrotron radiation. This evolution could align with models of GRB jets where the early photospheric emission fades, allowing synchrotron emission from internal shocks to become more prominent \citep{2000ApJ...530..292M}. In Region 2 of GRB 230307A, the low PF and softening $\alpha$ might reflect an intermediate phase where the jet's magnetic field is still restructuring, possibly initiating the stage for the changes (dominance of Synchrotron emission) in Region 3.

\subsection{Decay phase: transition to synchrotron emission}

In the third temporal bin (19–38 s), GRB 230307A exhibits a significant increase in polarization fraction (PF $>$ 49\%, 1 $\sigma$), and a further softened spectral index ($\alpha$ $\sim$ -1.49). The hint of high PF indicates the emergence of a more ordered magnetic field, a hallmark of synchrotron emission from a coherent field structure, as the tangled field from earlier phases aligns, possibly due to shock compression or jet expansion \citep{2009ApJ...698.1042T}. The soft $\alpha$, well below the slow-cooling synchrotron limit, suggests that the emission might be dominated by synchrotron radiation from a cooled electron population \citep{2011ApJ...726...90Z, 2020MNRAS.491.3343G}, potentially in a late internal shock scenario.
For GRB 230307A, recent spectral correlation analyses have suggested that late-phase emission may involve synchrotron radiation from ordered fields, with spectral softening reflecting the cooling of electrons \citep{2023ApJ...953L...8W}. The observed polarization properties of GRB 230307A in the decaying phase (high PF) are possibly consistent with this picture, providing suggesting a transition from a thermal-dominated (region 1) to a non-thermal-dominated (region 3) radiation mechanism.

The initial low polarization fraction in GRB 230307A suggests that the jet's magnetization might be relatively weak, characterized by a magnetization parameter $\sigma$ $<$ 1, indicating that the energy is primarily carried by the kinetic motion of particles rather than the magnetic field. In contrast, the high polarization fraction observed in the later, decaying phase points to a more magnetized jet with $\sigma$ $>$ 1, where magnetic energy dominates the jet dynamics \citep{2009ApJ...700L..65Z, 2020MNRAS.491.3343G}. This possible evolution aligns with theoretical models of magnetic jet acceleration, which predict that as the jet propagates outward, magnetic reconnection or dissipation processes can convert magnetic energy into particle acceleration, leading to a more ordered magnetic field and higher polarization \citep{2007MNRAS.380...51K, 2011MNRAS.418L..79T}. Additionally, the spectro-polarimetric evolution of GRB 230307A supports theoretical predictions of a transition from a thermally dominated emission (e.g., photospheric) to a non-thermal synchrotron-dominated regime, as evidenced by the spectral index softening from $\alpha$ $\sim$ -0.30 to $\alpha$ $\sim$ -1.85. Similar transitions have been observed in other bright GRBs, such as GRB 150309A, GRB 160625B, GRB 210619B, and GRB 230204B, where broadband spectral analyses revealed shifts in the dominant emission mechanism over time \citep{2024A&A...683A..55C, 2018NatAs...2...69Z, 2021ApJ...920...53C, 2023MNRAS.519.3201C}. Typically, such transitions are reported in multi-pulsed GRBs with collapsar origin; for instance, GRB 160325A exhibited a hard $\alpha$ and low PF in its first pulse, followed by a softer $\alpha$ and higher PF in its second pulse, mirroring the trend seen in GRB 230307A \citep{Sharma20, 2024ApJ...972..166G}. However, our analysis indicates GRB 230307A as a rare case where this transition occurs within a single emission episode associated with merger origin.

\section{Summary}\label{summary}

In this paper, we present a time-resolved spectro-polarimetric analysis of GRB 230307A, the second-brightest long gamma-ray burst ever observed, uniquely linked to a Kilonova \citep{2024Natur.626..737L}. Notably, GRB 230307A is the brightest GRB for which such detailed spectro-polarimetric analysis is performed. Using data from \astr CZTI, \frm GBM, and \kw, we uncover a hint of transition in the burst's emission mechanism: from an early phase dominated by thermal, photospheric emission (low polarization, hard spectral index) to a later phase dominated by non-thermal synchrotron emission (high polarization, softer spectral index). This plausible evolution, observed within a single emission episode, highlights the dynamic role of magnetic fields in shaping GRB jets and provides insights into the central engine's properties. Our findings underscore the importance of time-resolved spectro-polarimetric studies in understanding GRB physics.

Future observations of similar energetic GRBs, particularly with next-generation X-ray and gamma-ray polarimeters like COSI and POLAR-2, will be essential to further constrain the radiation mechanisms of GRBs. The ability to perform detailed time-resolved spectro-polarimetric analyses will provide new insights into the role of magnetic fields in GRB emission and help unravel the complex physics of these extraordinary cosmic events.

\begin{acknowledgements}
The authors acknowledge the referee for the valuable suggestions. SG acknowledges Prof A.R. Rao, Dr Mithun, Dr. Vidushi Sharma, and Dr. Shabnam Iyyani for their valuable comments. RG is thankful to Prof. Peter Veres, Dr. Ramandeep Gill, and Dr. Binbin Zhang for the fruitful discussion. This publication uses data from the AstroSat mission of the Indian Space Research Organisation (ISRO), archived at the Indian Space Science Data Centre (ISSDC). CZT-Imager is built by a consortium of institutes across India, including the Tata Institute of Fundamental Research (TIFR), Mumbai, the Vikram Sarabhai Space Centre, Thiruvananthapuram, the ISRO Satellite Centre (ISAC), Bangalore, the Inter-University Centre for Astronomy and Astrophysics, Pune, the Physical Research Laboratory, Ahmadabad, Space Application Centre, Ahmadabad. The Geant4 simulations for this paper were performed using the HPC resources at IUCAA. This research also has used data obtained through the HEASARC Online Service, provided by the NASA-GSFC, in support of NASA High Energy Astrophysics Programs. RG was sponsored by the National Aeronautics and Space Administration (NASA) through a contract with ORAU. The views and conclusions contained in this document are those of the authors and should not be interpreted as representing the official policies, either expressed or implied, of the National Aeronautics and Space Administration (NASA) or the U.S. Government. The U.S. Government is authorized to reproduce and distribute reprints for Government purposes, notwithstanding any copyright notation herein. The work of DF, DS, AL, AR, AT, and MU was supported by the basic
funding program of the Io e Institute FFUG-2024-0002.
\end{acknowledgements}

\bibliography{GRB230307A}{}
\bibliographystyle{aa}

\begin{appendix}
\onecolumn

\section{Time-Resolved Spectro-Polarimetry Results}

\begin{figure}[!ht]
\centering

    \includegraphics[width = 0.45\textwidth]{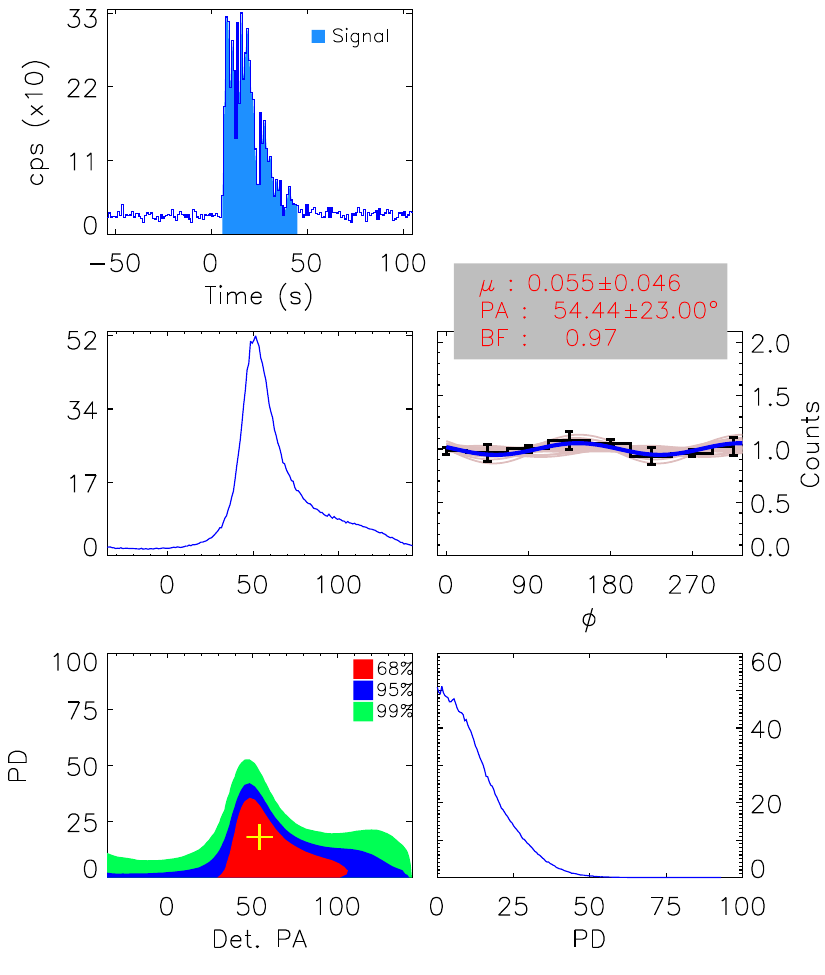}
    \includegraphics[width = 0.45\textwidth]{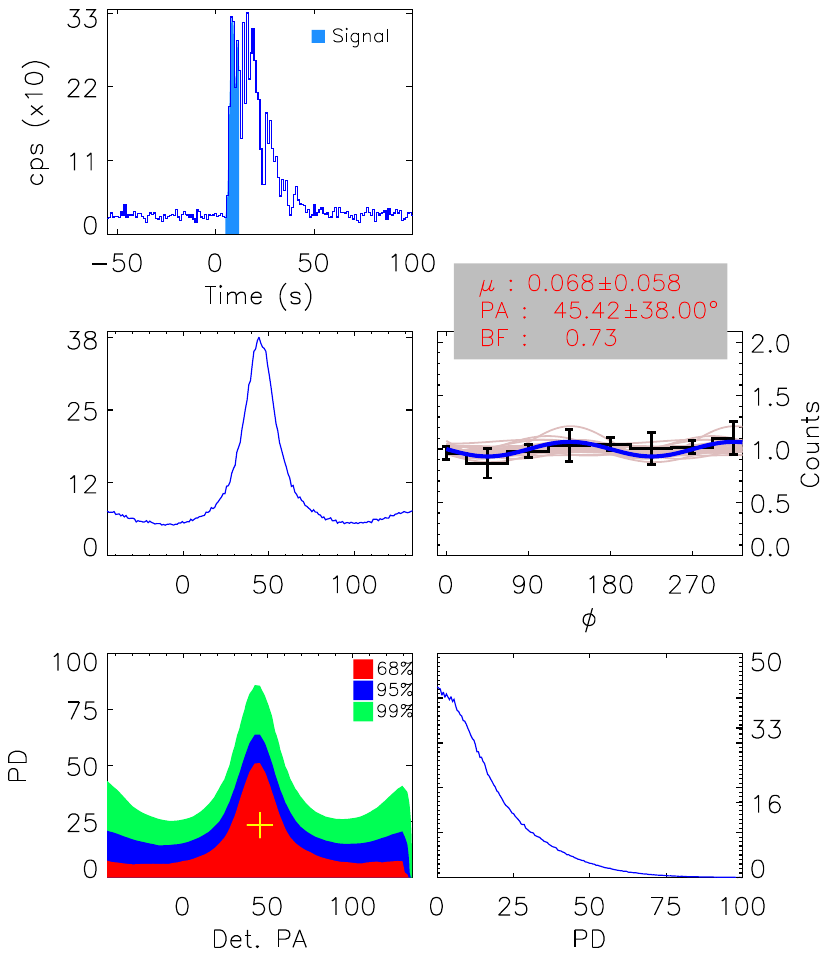}
    \includegraphics[width = 0.45\textwidth]{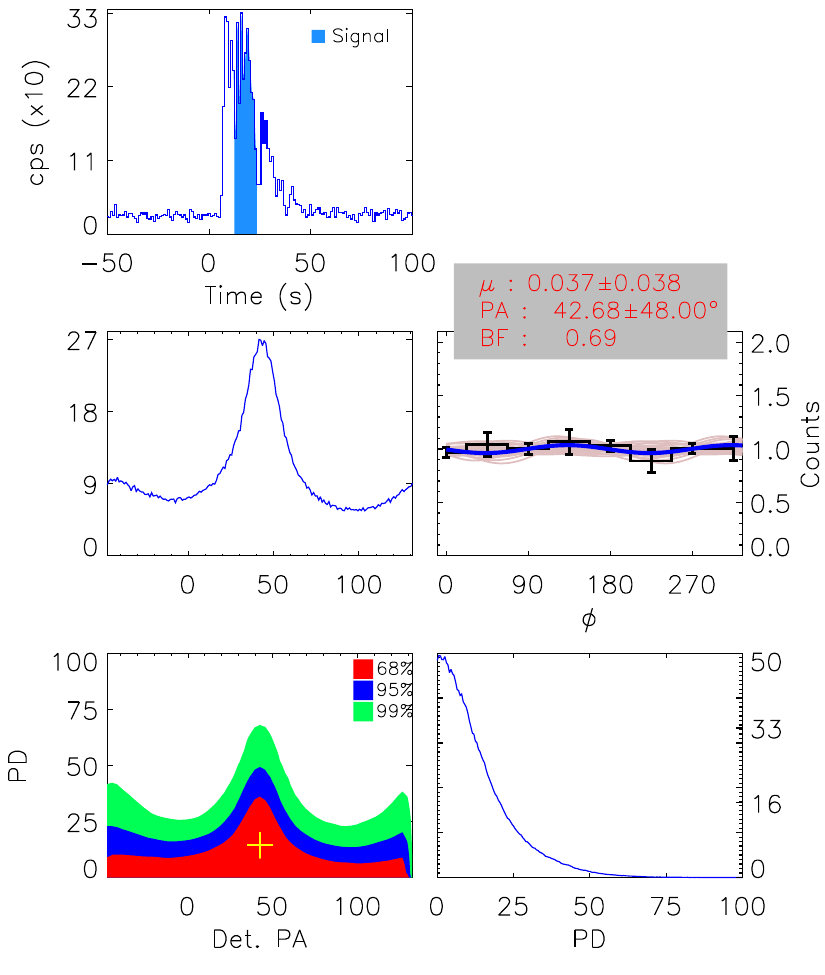}
    \includegraphics[width = 0.45\textwidth]{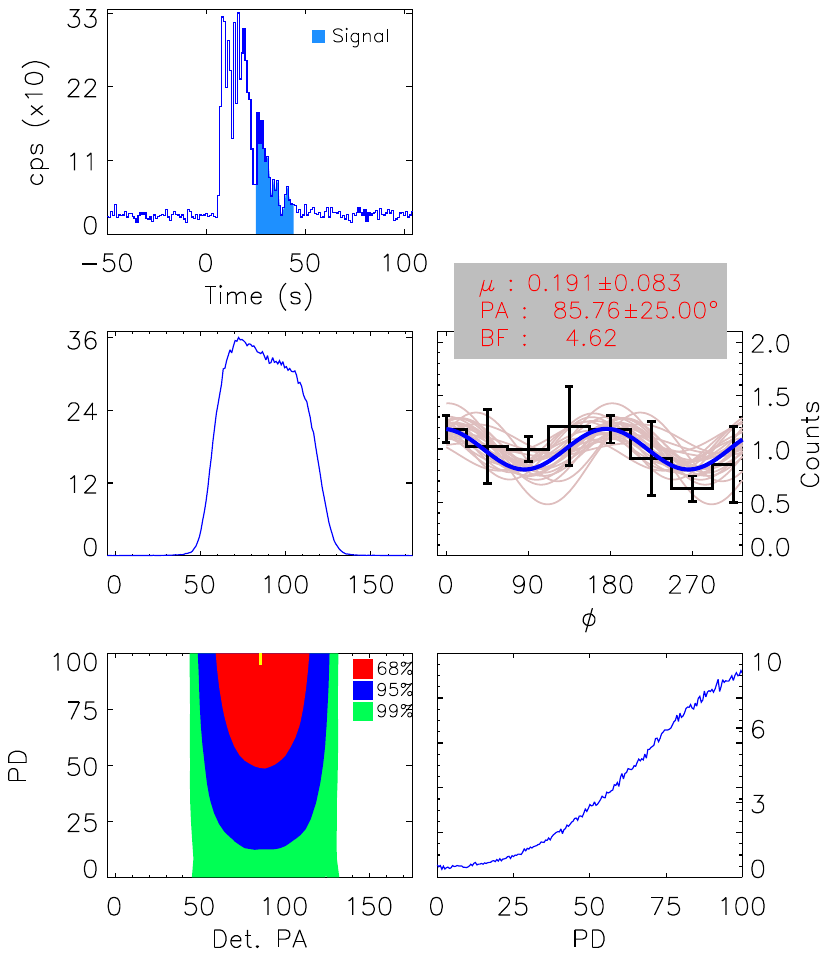}
    \caption{Time-integrated (top-left figure) and time-resolved (top-right figure for region 1, bottom-left figure for region 2, and bottom-right figure for region 3) polarization analysis of GRB 230307A in 100-600 keV using \astr CZTI. Each figure is composed of five panels: (1) the Compton light curve for the region of interest (time-integrated, region 1, region 2, and region 3), obtained using Bayesian block analysis (blue-shaded region, top panel); (2) the posterior probability distribution of the detector polarization angle (middle-left panel); (3) the modulation curve of the burst, illustrating the azimuthal dependence of the detected counts (middle-right panel) where in blue is the best fit curve and pink are the 250 randomly selected fitted modulation curves; (4) a two-dimensional contour plot depicting the joint posterior distribution of the detector polarization angle and polarization fraction (denoted as PD in the plots, equivalent to the PF referenced in the text; bottom-left panel); and (5) the posterior probability distribution of the polarization fraction (bottom-right panel). The Bayes factor, quantifying the strength of evidence for polarization, is computed for the 100–600 keV energy range.}
    \label{pol-res}
\end{figure}

\small
\begin{longtable}{cccccccc}
\caption{The \textit{Fermi} GBM time-resolved spectral analysis result for the GRB 230307A.}
\label{table_spec} \\
\hline\hline
$(T_{i},T_{f})$ &  $\alpha$ & $\beta$ & $E_p$ & $p$ & $E_{c}$& $\Delta BIC$&Flux\, (8-900\, keV)\\
(s) & & & (keV) & & (keV) & $(BIC_{Band}-BIC_{CPL})$& $\mathrm{erg\,cm^{-2}\,s^{-1}}$\\
\hline
\endfirsthead  

\caption{Continued.} \\  
\hline  
$(T_{i},T_{f})$ &  $\alpha$ & $\beta$ & $E_p$ & $p$ & $E_{c}$& $\Delta BIC$&Flux\, (8-900\, keV)\\
(s) & & & (keV) & & (keV) & $(BIC_{Band}-BIC_{CPL})$& $\mathrm{erg\,cm^{-2}\,s^{-1}}$\\
\hline  
\endhead  

\hline  
\endfoot
\hline
\endlastfoot
$(-0.02,0.12)$& $-0.51^{0.07}_{-0.06}$ & $-3.08^{0.16}_{-0.16}$ & $189.28^{8.77}_{-8.69}$ & $-0.69^{0.04}_{-0.04}$ & $168.00^{9.25}_{-8.97}$ & $-28.72$ & $-4.26^{0.01}_{-0.01}$\\
$(0.12,0.30)$& $-0.71^{0.08}_{-0.08}$ & $-3.48^{0.44}_{-0.23}$ & $163.78^{9.99}_{-9.63}$ & $-0.83^{0.04}_{-0.05}$ & $155.56^{9.17}_{-9.12}$ & $-17.24$ & $-4.47^{0.01}_{-0.02}$\\
$(0.45,0.56)$& $-1.50^{0.10}_{-0.10}$ & $-5.43^{1.80}_{-1.79}$ & $175.12^{26.73}_{-28.94}$ & $-1.50^{0.10}_{-0.10}$ & $371.84^{107.65}_{-114.38}$ & $-23.65$ & $-5.09^{0.03}_{-0.03}$\\
$(0.70,0.89)$& $-0.56^{0.03}_{-0.03}$ & $-6.21^{1.25}_{-1.25}$ & $652.30^{17.98}_{-17.64}$ & $-0.56^{0.03}_{-0.03}$ & $456.46^{18.56}_{-18.73}$ & $-3.22$ & $-3.92^{0.01}_{-0.01}$\\
$(0.89,0.97)$& $-0.49^{0.05}_{-0.05}$ & $-6.32^{1.18}_{-1.18}$ & $602.06^{24.25}_{-24.05}$ & $-0.49^{0.05}_{-0.04}$ & $398.91^{24.23}_{-24.79}$ & $-1.64$ & $-3.98^{0.01}_{-0.01}$\\
$(0.97,1.06)$& $-0.38^{0.04}_{-0.04}$ & $-5.66^{1.22}_{-1.41}$ & $677.58^{20.22}_{-19.78}$ & $-0.39^{0.04}_{-0.03}$ & $426.10^{19.39}_{-19.64}$ & $-1.21$ & $-3.73^{0.01}_{-0.01}$\\
$(1.06,1.28)$& $-0.60^{0.03}_{-0.03}$ & $-6.33^{1.18}_{-1.13}$ & $508.24^{13.12}_{-13.16}$ & $-0.60^{0.03}_{-0.03}$ & $365.10^{14.88}_{-14.68}$ & $-4.24$ & $-4.06^{0.01}_{-0.01}$\\
$(1.28,1.47)$& $-0.60^{0.03}_{-0.03}$ & $-4.83^{1.01}_{-1.35}$ & $616.58^{19.15}_{-18.92}$ & $-0.60^{0.03}_{-0.03}$ & $445.75^{20.54}_{-19.77}$ & $-8.44$ & $-4.00^{0.01}_{-0.01}$\\
$(1.47,1.53)$& $-0.27^{0.04}_{-0.04}$ & $-5.55^{1.11}_{-1.31}$ & $1107.60^{31.75}_{-31.32}$ & $-0.28^{0.03}_{-0.03}$ & $648.86^{25.90}_{-25.89}$ & $0.30$ & $-3.31^{0.01}_{-0.01}$\\
$(1.53,1.66)$& $-0.21^{0.02}_{-0.02}$ & $-6.93^{0.76}_{-0.75}$ & $993.86^{15.97}_{-15.83}$ & $-0.22^{0.02}_{-0.02}$ & $557.98^{14.31}_{-14.06}$ & $3.72$ & $-3.32^{0.01}_{-0.01}$\\
$(1.66,1.73)$& $-0.31^{0.03}_{-0.03}$ & $-6.74^{0.90}_{-0.89}$ & $1265.60^{32.28}_{-32.41}$ & $-0.31^{0.03}_{-0.03}$ & $748.97^{29.11}_{-28.44}$ & $2.09$ & $-3.31^{0.01}_{-0.01}$\\
$(1.73,1.80)$& $-0.31^{0.03}_{-0.03}$ & $-6.48^{0.96}_{-0.99}$ & $1355.20^{33.62}_{-33.42}$ & $-0.32^{0.03}_{-0.03}$ & $806.41^{29.79}_{-29.97}$ & $1.28$ & $-3.27^{0.01}_{-0.01}$\\
$(1.80,2.00)$& $-0.17^{0.02}_{-0.02}$ & $-6.51^{0.87}_{-0.94}$ & $1249.15^{16.32}_{-16.00}$ & $-0.18^{0.02}_{-0.02}$ & $687.23^{13.85}_{-13.61}$ & $3.86$ & $-3.20^{0.00}_{-0.01}$\\
$(2.00,2.33)$& $-0.20^{0.02}_{-0.02}$ & $-6.70^{0.83}_{-0.86}$ & $785.43^{9.47}_{-9.45}$ & $-0.21^{0.02}_{-0.02}$ & $438.28^{8.46}_{-8.36}$ & $3.68$ & $-3.56^{0.00}_{0.00}$\\
$(2.33,2.39)$& $-0.50^{0.05}_{-0.05}$ & $-5.86^{1.47}_{-1.48}$ & $707.34^{37.16}_{-35.75}$ & $-0.51^{0.05}_{-0.05}$ & $483.46^{37.20}_{-36.62}$ & $-2.03$ & $-3.89^{0.02}_{-0.02}$\\
$(2.39,2.67)$& $-0.18^{0.02}_{-0.02}$ & $-7.02^{0.75}_{-0.72}$ & $731.45^{9.66}_{-9.67}$ & $-0.18^{0.02}_{-0.02}$ & $402.54^{9.06}_{-8.97}$ & $5.19$ & $-3.63^{0.00}_{-0.01}$\\
$(2.67,2.72)$& $-0.26^{0.06}_{-0.06}$ & $-5.97^{1.22}_{-1.32}$ & $645.32^{25.40}_{-25.96}$ & $-0.26^{0.06}_{-0.06}$ & $374.52^{24.84}_{-24.56}$ & $2.61$ & $-3.82^{0.01}_{-0.01}$\\
$(2.72,2.84)$& $-0.31^{0.04}_{-0.04}$ & $-6.60^{1.04}_{-1.00}$ & $499.38^{13.65}_{-13.77}$ & $-0.31^{0.04}_{-0.04}$ & $296.59^{14.66}_{-14.54}$ & $2.03$ & $-4.01^{0.01}_{-0.01}$\\
$(2.84,2.89)$& $-0.46^{0.07}_{-0.07}$ & $-5.51^{1.46}_{-1.62}$ & $452.41^{24.07}_{-23.21}$ & $-0.47^{0.07}_{-0.07}$ & $298.88^{25.59}_{-25.65}$ & $-2.04$ & $-4.12^{0.02}_{-0.02}$\\
$(2.89,3.06)$& $-0.69^{0.05}_{-0.05}$ & $-5.48^{1.50}_{-1.58}$ & $443.98^{20.63}_{-21.00}$ & $-0.69^{0.04}_{-0.04}$ & $342.28^{24.07}_{-24.11}$ & $-6.72$ & $-4.35^{0.01}_{-0.01}$\\
$(7.06,7.13)$& $-0.66^{0.03}_{-0.03}$ & $-5.85^{1.36}_{-1.44}$ & $858.58^{33.72}_{-33.49}$ & $-0.67^{0.03}_{-0.03}$ & $649.48^{34.81}_{-35.46}$ & $-5.94$ & $-3.67^{0.01}_{-0.01}$\\
$(7.13,7.20)$& $-0.72^{0.04}_{-0.04}$ & $-5.17^{1.25}_{-1.51}$ & $664.64^{32.33}_{-32.02}$ & $-0.73^{0.04}_{-0.04}$ & $531.21^{37.43}_{-37.58}$ & $-9.40$ & $-3.91^{0.01}_{-0.01}$\\
$(7.20,7.30)$& $-0.59^{0.02}_{-0.02}$ & $-6.11^{1.22}_{-1.25}$ & $865.80^{23.81}_{-23.63}$ & $-0.60^{0.02}_{-0.02}$ & $620.10^{22.77}_{-22.65}$ & $-4.41$ & $-3.53^{0.01}_{-0.01}$\\
$(7.30,7.49)$& $-0.71^{0.02}_{-0.02}$ & $-6.14^{1.23}_{-1.26}$ & $757.39^{17.37}_{-17.51}$ & $-0.71^{0.02}_{-0.02}$ & $589.24^{19.63}_{-19.93}$ & $-6.61$ & $-3.70^{0.01}_{-0.01}$\\
$(7.49,7.69)$& $-0.73^{0.03}_{-0.03}$ & $-5.40^{1.36}_{-1.70}$ & $452.52^{13.23}_{-13.70}$ & $-0.74^{0.03}_{-0.03}$ & $365.05^{16.28}_{-16.37}$ & $-8.99$ & $-4.02^{0.01}_{-0.01}$\\
$(7.69,7.98)$& $-0.73^{0.01}_{-0.01}$ & $-6.27^{1.01}_{-1.12}$ & $772.74^{13.70}_{-14.08}$ & $-0.73^{0.01}_{-0.01}$ & $612.44^{15.95}_{-15.68}$ & $-7.77$ & $-3.66^{0.01}_{-0.01}$\\
$(7.98,8.15)$& $-0.62^{0.02}_{-0.02}$ & $-6.78^{0.89}_{-0.87}$ & $968.97^{18.00}_{-17.78}$ & $-0.62^{0.02}_{-0.02}$ & $701.99^{18.36}_{-18.88}$ & $-4.57$ & $-3.42^{0.01}_{-0.01}$\\
$(8.15,8.34)$& $-0.65^{0.01}_{-0.01}$ & $-5.76^{1.07}_{-1.23}$ & $1268.56^{24.14}_{-24.58}$ & $-0.65^{0.01}_{-0.01}$ & $949.89^{23.10}_{-23.16}$ & $-7.49$ & $-3.30^{0.01}_{-0.01}$\\
$(8.34,8.42)$& $-0.63^{0.02}_{-0.02}$ & $-5.17^{1.07}_{-1.49}$ & $1092.28^{36.67}_{-36.60}$ & $-0.64^{0.02}_{-0.02}$ & $820.14^{32.39}_{-32.65}$ & $-8.22$ & $-3.42^{0.01}_{-0.01}$\\
$(8.42,8.62)$& $-0.64^{0.01}_{-0.01}$ & $-6.46^{0.99}_{-1.03}$ & $1409.12^{24.04}_{-24.17}$ & $-0.64^{0.01}_{-0.01}$ & $1038.96^{23.60}_{-23.53}$ & $-5.38$ & $-3.25^{0.01}_{-0.01}$\\
$(8.62,8.92)$& $-0.67^{0.01}_{-0.01}$ & $-7.69^{1.54}_{-1.54}$ & $918.53^{15.49}_{-15.27}$ & $-0.67^{0.01}_{-0.01}$ & $692.34^{16.89}_{-16.56}$ & $-5.89$ & $-3.58^{0.01}_{-0.01}$\\
$(8.92,8.98)$& $-0.69^{0.03}_{-0.03}$ & $-6.90^{2.15}_{-2.12}$ & $859.61^{36.70}_{-37.16}$ & $-0.70^{0.03}_{-0.03}$ & $665.27^{39.95}_{-40.45}$ & $-6.18$ & $-3.71^{0.01}_{-0.01}$\\
$(8.98,9.10)$& $-0.64^{0.02}_{-0.02}$ & $-7.33^{1.79}_{-1.83}$ & $840.15^{22.09}_{-21.98}$ & $-0.64^{0.02}_{-0.02}$ & $620.70^{24.84}_{-24.38}$ & $-5.12$ & $-3.63^{0.01}_{-0.01}$\\
$(9.10,9.21)$& $-0.75^{0.03}_{-0.03}$ & $-7.15^{1.86}_{-1.96}$ & $693.49^{23.48}_{-23.88}$ & $-0.75^{0.03}_{-0.03}$ & $555.36^{29.36}_{-28.88}$ & $-7.44$ & $-3.82^{0.01}_{-0.01}$\\
$(9.21,9.31)$& $-0.77^{0.02}_{-0.02}$ & $-5.63^{1.60}_{-2.08}$ & $811.83^{27.57}_{-27.90}$ & $-0.77^{0.02}_{-0.02}$ & $667.18^{30.61}_{-30.61}$ & $-11.34$ & $-3.70^{0.01}_{-0.01}$\\
$(9.31,9.53)$& $-0.69^{0.02}_{-0.02}$ & $-7.55^{1.57}_{-1.64}$ & $793.06^{14.77}_{-15.05}$ & $-0.69^{0.01}_{-0.02}$ & $606.18^{16.52}_{-16.63}$ & $-6.25$ & $-3.59^{0.01}_{-0.01}$\\
$(9.53,9.70)$& $-0.77^{0.02}_{-0.02}$ & $-7.53^{1.82}_{-1.74}$ & $1062.91^{27.88}_{-27.64}$ & $-0.77^{0.02}_{-0.02}$ & $865.60^{29.19}_{-30.05}$ & $-7.78$ & $-3.59^{0.01}_{-0.01}$\\
$(9.70,9.79)$& $-0.72^{0.03}_{-0.03}$ & $-7.53^{1.77}_{-1.74}$ & $967.73^{35.62}_{-36.14}$ & $-0.72^{0.03}_{-0.03}$ & $756.92^{40.87}_{-40.41}$ & $-6.65$ & $-3.69^{0.01}_{-0.01}$\\
$(9.79,9.89)$& $-0.71^{0.02}_{-0.02}$ & $-5.89^{1.71}_{-2.29}$ & $1260.39^{38.35}_{-38.18}$ & $-0.72^{0.02}_{-0.02}$ & $988.31^{40.47}_{-40.80}$ & $-9.86$ & $-3.50^{0.01}_{-0.01}$\\
$(9.89,9.99)$& $-0.75^{0.03}_{-0.03}$ & $-7.54^{1.66}_{-1.64}$ & $882.06^{32.16}_{-32.90}$ & $-0.75^{0.03}_{-0.03}$ & $705.25^{38.64}_{-37.60}$ & $-7.36$ & $-3.75^{0.01}_{-0.01}$\\
$(9.99,10.11)$& $-0.71^{0.02}_{-0.02}$ & $-7.48^{1.72}_{-1.71}$ & $1125.06^{28.86}_{-29.64}$ & $-0.71^{0.02}_{-0.02}$ & $873.84^{31.10}_{-32.03}$ & $-6.54$ & $-3.50^{0.01}_{-0.01}$\\
$(10.11,10.33)$& $-0.68^{0.01}_{-0.01}$ & $-5.75^{1.28}_{-1.73}$ & $1134.52^{21.63}_{-22.14}$ & $-0.68^{0.01}_{-0.01}$ & $870.73^{22.32}_{-22.57}$ & $-10.49$ & $-3.42^{0.01}_{-0.01}$\\
$(10.33,10.54)$& $-0.70^{0.02}_{-0.02}$ & $-6.47^{1.90}_{-2.20}$ & $880.47^{21.18}_{-20.67}$ & $-0.70^{0.02}_{-0.02}$ & $681.47^{21.64}_{-21.83}$ & $-7.98$ & $-3.66^{0.01}_{-0.01}$\\
$(10.54,10.65)$& $-0.72^{0.02}_{-0.02}$ & $-7.54^{1.67}_{-1.69}$ & $1003.66^{25.90}_{-26.34}$ & $-0.72^{0.02}_{-0.02}$ & $785.49^{28.92}_{-28.47}$ & $-6.77$ & $-3.48^{0.01}_{-0.01}$\\
$(10.65,10.99)$& $-0.77^{0.01}_{-0.01}$ & $-7.60^{1.57}_{-1.61}$ & $824.50^{15.22}_{-15.46}$ & $-0.77^{0.01}_{-0.01}$ & $673.30^{18.43}_{-18.38}$ & $-7.95$ & $-3.71^{0.01}_{-0.01}$\\
$(10.99,11.19)$& $-0.80^{0.02}_{-0.02}$ & $-7.45^{1.71}_{-1.70}$ & $681.58^{18.76}_{-18.78}$ & $-0.80^{0.02}_{-0.02}$ & $569.94^{23.35}_{-23.69}$ & $-8.54$ & $-3.90^{0.01}_{-0.01}$\\
$(11.19,11.41)$& $-0.80^{0.02}_{-0.02}$ & $-7.11^{1.78}_{-1.89}$ & $864.35^{20.21}_{-20.49}$ & $-0.80^{0.02}_{-0.02}$ & $724.48^{23.13}_{-23.47}$ & $-9.14$ & $-3.66^{0.01}_{-0.01}$\\
$(11.41,11.75)$& $-0.88^{0.02}_{-0.02}$ & $-6.86^{1.90}_{-2.03}$ & $624.99^{14.65}_{-14.37}$ & $-0.88^{0.02}_{-0.02}$ & $559.81^{19.60}_{-20.04}$ & $-10.93$ & $-3.92^{0.01}_{-0.01}$\\
$(11.75,11.82)$& $-0.97^{0.04}_{-0.04}$ & $-7.22^{1.96}_{-1.93}$ & $522.08^{31.27}_{-31.66}$ & $-0.96^{0.04}_{-0.04}$ & $506.37^{45.80}_{-46.38}$ & $-11.99$ & $-4.13^{0.02}_{-0.02}$\\
$(11.82,11.97)$& $-0.87^{0.02}_{-0.02}$ & $-7.19^{1.79}_{-1.86}$ & $1027.32^{30.62}_{-29.49}$ & $-0.87^{0.02}_{-0.02}$ & $914.75^{36.30}_{-35.97}$ & $-10.34$ & $-3.62^{0.01}_{-0.01}$\\
$(11.97,12.10)$& $-0.86^{0.02}_{-0.02}$ & $-7.58^{1.69}_{-1.72}$ & $822.56^{26.44}_{-25.74}$ & $-0.86^{0.02}_{-0.02}$ & $725.80^{33.46}_{-32.49}$ & $-9.83$ & $-3.73^{0.01}_{-0.01}$\\
$(12.10,12.41)$& $-0.91^{0.01}_{-0.01}$ & $-7.24^{1.73}_{-1.88}$ & $863.29^{20.60}_{-20.72}$ & $-0.91^{0.01}_{-0.01}$ & $792.65^{26.63}_{-26.90}$ & $-11.10$ & $-3.81^{0.01}_{-0.01}$\\
$(12.41,12.53)$& $-0.91^{0.02}_{-0.02}$ & $-6.69^{2.14}_{-2.27}$ & $952.68^{37.25}_{-36.06}$ & $-0.92^{0.02}_{-0.02}$ & $884.75^{47.09}_{-46.13}$ & $-11.42$ & $-3.72^{0.01}_{-0.01}$\\
$(12.53,12.63)$& $-0.94^{0.02}_{-0.02}$ & $-7.12^{1.93}_{-2.00}$ & $1031.01^{46.61}_{-45.89}$ & $-0.94^{0.02}_{-0.02}$ & $970.81^{57.16}_{-60.04}$ & $-11.52$ & $-3.79^{0.01}_{-0.01}$\\
$(12.63,12.70)$& $-0.80^{0.03}_{-0.03}$ & $-7.28^{1.93}_{-1.91}$ & $970.38^{39.53}_{-39.37}$ & $-0.80^{0.03}_{-0.03}$ & $812.88^{46.68}_{-47.79}$ & $-8.51$ & $-3.63^{0.01}_{-0.01}$\\
$(12.70,12.90)$& $-0.85^{0.02}_{-0.02}$ & $-7.07^{1.96}_{-2.02}$ & $723.80^{19.95}_{-19.36}$ & $-0.85^{0.02}_{-0.02}$ & $633.32^{24.47}_{-24.06}$ & $-9.82$ & $-3.79^{0.01}_{-0.01}$\\
$(12.90,13.03)$& $-0.86^{0.02}_{-0.02}$ & $-4.82^{1.00}_{-0.99}$ & $758.79^{25.35}_{-25.98}$ & $-0.88^{0.02}_{-0.02}$ & $690.84^{28.74}_{-28.78}$ & $-16.45$ & $-3.61^{0.01}_{-0.01}$\\
$(13.03,13.37)$& $-0.95^{0.01}_{-0.01}$ & $-7.33^{1.76}_{-1.78}$ & $874.74^{17.41}_{-18.06}$ & $-0.95^{0.01}_{-0.01}$ & $834.77^{22.10}_{-23.06}$ & $-11.98$ & $-3.66^{0.01}_{-0.01}$\\
$(13.37,13.49)$& $-1.00^{0.02}_{-0.02}$ & $-7.02^{1.97}_{-2.02}$ & $867.58^{33.08}_{-33.46}$ & $-1.00^{0.02}_{-0.02}$ & $870.14^{46.19}_{-46.88}$ & $-12.85$ & $-3.77^{0.01}_{-0.01}$\\
$(13.49,13.95)$& $-1.03^{0.01}_{-0.01}$ & $-7.26^{1.91}_{-1.88}$ & $731.91^{17.98}_{-19.25}$ & $-1.03^{0.01}_{-0.01}$ & $755.78^{26.44}_{-27.27}$ & $-13.53$ & $-3.98^{0.01}_{-0.01}$\\
$(13.95,14.25)$& $-1.03^{0.01}_{-0.01}$ & $-6.58^{1.99}_{-2.25}$ & $845.26^{25.42}_{-25.33}$ & $-1.03^{0.01}_{-0.01}$ & $879.79^{36.19}_{-36.60}$ & $-14.96$ & $-3.87^{0.01}_{-0.01}$\\
$(14.25,14.40)$& $-1.01^{0.02}_{-0.02}$ & $-7.43^{1.71}_{-1.75}$ & $969.89^{34.19}_{-33.02}$ & $-1.01^{0.02}_{-0.02}$ & $985.59^{46.93}_{-47.37}$ & $-13.21$ & $-3.75^{0.01}_{-0.01}$\\
$(14.40,14.48)$& $-1.07^{0.05}_{-0.05}$ & $-6.55^{2.39}_{-2.38}$ & $432.88^{32.25}_{-32.73}$ & $-1.07^{0.05}_{-0.05}$ & $470.74^{53.71}_{-54.06}$ & $-14.19$ & $-4.29^{0.02}_{-0.02}$\\
$(14.48,14.54)$& $-1.03^{0.05}_{-0.05}$ & $-6.51^{2.32}_{-2.34}$ & $485.59^{35.65}_{-34.71}$ & $-1.03^{0.05}_{-0.05}$ & $504.59^{54.43}_{-54.74}$ & $-13.39$ & $-4.14^{0.02}_{-0.02}$\\
$(14.54,14.75)$& $-1.14^{0.02}_{-0.02}$ & $-6.93^{2.10}_{-2.13}$ & $596.10^{24.16}_{-23.81}$ & $-1.14^{0.02}_{-0.02}$ & $694.46^{40.13}_{-38.94}$ & $-16.05$ & $-4.02^{0.01}_{-0.01}$\\
$(14.75,14.97)$& $-1.15^{0.02}_{-0.02}$ & $-7.07^{2.00}_{-2.05}$ & $638.21^{27.68}_{-28.18}$ & $-1.15^{0.02}_{-0.02}$ & $750.38^{45.25}_{-45.54}$ & $-16.24$ & $-4.07^{0.01}_{-0.01}$\\
$(14.97,15.12)$& $-1.01^{0.02}_{-0.02}$ & $-6.94^{1.96}_{-1.99}$ & $734.37^{25.66}_{-26.08}$ & $-1.01^{0.02}_{-0.02}$ & $744.88^{35.57}_{-36.02}$ & $-13.30$ & $-3.81^{0.01}_{-0.01}$\\
$(15.12,15.37)$& $-1.04^{0.02}_{-0.02}$ & $-6.18^{2.10}_{-2.42}$ & $556.90^{19.80}_{-19.72}$ & $-1.04^{0.02}_{-0.02}$ & $587.40^{29.51}_{-29.30}$ & $-15.38$ & $-4.04^{0.01}_{-0.01}$\\
$(15.37,15.47)$& $-1.12^{0.03}_{-0.03}$ & $-7.12^{1.98}_{-1.97}$ & $588.96^{36.76}_{-38.31}$ & $-1.12^{0.03}_{-0.03}$ & $673.22^{61.56}_{-62.28}$ & $-15.64$ & $-4.13^{0.01}_{-0.01}$\\
$(15.47,15.78)$& $-1.09^{0.02}_{-0.02}$ & $-7.13^{1.92}_{-1.93}$ & $681.39^{22.71}_{-22.87}$ & $-1.09^{0.02}_{-0.02}$ & $752.36^{34.03}_{-34.29}$ & $-15.02$ & $-4.01^{0.01}_{-0.01}$\\
$(15.78,16.05)$& $-1.13^{0.02}_{-0.02}$ & $-6.75^{2.12}_{-2.20}$ & $840.79^{33.54}_{-34.31}$ & $-1.13^{0.02}_{-0.02}$ & $970.07^{49.65}_{-51.24}$ & $-16.41$ & $-4.02^{0.01}_{-0.01}$\\
$(16.05,16.35)$& $-1.06^{0.02}_{-0.02}$ & $-6.87^{2.07}_{-2.09}$ & $815.17^{30.39}_{-30.31}$ & $-1.06^{0.02}_{-0.02}$ & $875.20^{44.45}_{-44.34}$ & $-14.50$ & $-4.06^{0.01}_{-0.01}$\\
$(16.35,16.66)$& $-1.06^{0.02}_{-0.02}$ & $-5.02^{1.49}_{-2.36}$ & $717.62^{29.19}_{-29.40}$ & $-1.07^{0.02}_{-0.02}$ & $786.60^{39.64}_{-39.36}$ & $-19.24$ & $-4.04^{0.01}_{-0.01}$\\
$(16.66,17.42)$& $-1.15^{0.01}_{-0.01}$ & $-7.58^{1.63}_{-1.68}$ & $548.04^{15.15}_{-15.03}$ & $-1.15^{0.01}_{-0.01}$ & $648.03^{25.36}_{-25.99}$ & $-16.40$ & $-4.26^{0.01}_{-0.01}$\\
$(17.42,17.61)$& $-1.09^{0.03}_{-0.03}$ & $-6.43^{2.24}_{-2.36}$ & $484.72^{26.25}_{-27.23}$ & $-1.10^{0.03}_{-0.03}$ & $538.61^{44.89}_{-44.85}$ & $-15.67$ & $-4.33^{0.01}_{-0.01}$\\
$(17.61,17.88)$& $-1.22^{0.04}_{-0.04}$ & $-6.17^{2.63}_{-2.66}$ & $277.20^{17.46}_{-17.92}$ & $-1.23^{0.04}_{-0.04}$ & $363.38^{36.37}_{-37.26}$ & $-18.94$ & $-4.69^{0.02}_{-0.01}$\\
$(17.88,18.10)$& $-1.28^{0.07}_{-0.07}$ & $-6.62^{2.23}_{-2.24}$ & $178.53^{13.04}_{-13.68}$ & $-1.28^{0.07}_{-0.07}$ & $252.52^{35.55}_{-35.63}$ & $-19.06$ & $-4.99^{0.02}_{-0.02}$\\
$(18.10,18.52)$& $-1.61^{0.06}_{-0.06}$ & $-6.13^{2.65}_{-2.66}$ & $238.40^{33.73}_{-36.88}$ & $-1.61^{0.06}_{-0.06}$ & $643.15^{174.22}_{-168.16}$ & $-27.84$ & $-5.12^{0.03}_{-0.02}$\\
$(18.52,18.74)$& $-1.47^{0.05}_{-0.05}$ & $-6.63^{2.26}_{-2.33}$ & $330.73^{43.55}_{-43.97}$ & $-1.47^{0.05}_{-0.05}$ & $635.71^{128.82}_{-133.15}$ & $-23.63$ & $-4.88^{0.02}_{-0.02}$\\
$(18.74,18.99)$& $-1.37^{0.03}_{-0.04}$ & $-6.86^{2.21}_{-2.13}$ & $388.39^{32.22}_{-32.12}$ & $-1.37^{0.04}_{-0.04}$ & $619.07^{76.20}_{-78.72}$ & $-21.33$ & $-4.64^{0.02}_{-0.02}$\\
$(18.99,19.09)$& $-1.21^{0.07}_{-0.07}$ & $-4.82^{2.18}_{-3.11}$ & $348.57^{48.71}_{-52.43}$ & $-1.25^{0.05}_{-0.05}$ & $507.68^{80.15}_{-82.11}$ & $-22.64$ & $-4.48^{0.05}_{-0.03}$\\
$(19.09,19.88)$& $-1.23^{0.01}_{-0.01}$ & $-7.34^{1.89}_{-1.86}$ & $579.99^{17.00}_{-17.29}$ & $-1.23^{0.01}_{-0.01}$ & $756.02^{31.02}_{-31.06}$ & $-18.24$ & $-4.22^{0.01}_{-0.01}$\\
$(19.88,20.06)$& $-1.20^{0.02}_{-0.02}$ & $-6.94^{2.11}_{-2.09}$ & $742.63^{38.15}_{-39.13}$ & $-1.20^{0.02}_{-0.02}$ & $935.84^{65.70}_{-65.06}$ & $-17.53$ & $-4.02^{0.01}_{-0.01}$\\
$(20.06,20.40)$& $-1.18^{0.01}_{-0.02}$ & $-7.05^{2.03}_{-2.03}$ & $861.41^{36.12}_{-35.93}$ & $-1.18^{0.02}_{-0.02}$ & $1056.16^{57.16}_{-55.58}$ & $-17.02$ & $-4.06^{0.01}_{-0.01}$\\
$(20.40,21.15)$& $-1.14^{0.01}_{-0.01}$ & $-7.28^{1.86}_{-1.85}$ & $615.36^{17.95}_{-17.87}$ & $-1.14^{0.01}_{-0.01}$ & $718.95^{30.18}_{-30.58}$ & $-16.16$ & $-4.25^{0.01}_{-0.01}$\\
$(21.15,21.55)$& $-1.13^{0.02}_{-0.02}$ & $-6.08^{2.13}_{-2.56}$ & $643.10^{23.32}_{-23.49}$ & $-1.13^{0.02}_{-0.02}$ & $750.19^{36.09}_{-35.75}$ & $-18.06$ & $-4.09^{0.01}_{-0.01}$\\
$(21.55,21.68)$& $-1.10^{0.03}_{-0.03}$ & $-5.87^{2.22}_{-2.54}$ & $587.43^{37.70}_{-37.50}$ & $-1.11^{0.03}_{-0.03}$ & $668.48^{62.16}_{-62.42}$ & $-17.27$ & $-4.17^{0.02}_{-0.02}$\\
$(21.68,22.07)$& $-1.17^{0.02}_{-0.02}$ & $-7.22^{1.88}_{-1.88}$ & $591.36^{25.61}_{-25.09}$ & $-1.17^{0.02}_{-0.02}$ & $713.02^{44.95}_{-43.94}$ & $-16.73$ & $-4.29^{0.01}_{-0.01}$\\
$(22.07,22.15)$& $-1.18^{0.04}_{-0.04}$ & $-6.08^{2.66}_{-2.65}$ & $621.63^{61.50}_{-55.60}$ & $-1.19^{0.04}_{-0.04}$ & $782.16^{93.74}_{-94.03}$ & $-17.52$ & $-4.19^{0.03}_{-0.02}$\\
$(22.15,23.23)$& $-1.17^{0.01}_{-0.01}$ & $-7.35^{1.86}_{-1.85}$ & $494.95^{13.10}_{-12.99}$ & $-1.18^{0.01}_{-0.01}$ & $601.09^{23.59}_{-23.94}$ & $-16.90$ & $-4.36^{0.01}_{-0.01}$\\
$(23.23,23.52)$& $-1.17^{0.04}_{-0.04}$ & $-6.66^{2.20}_{-2.27}$ & $242.67^{11.58}_{-11.10}$ & $-1.17^{0.04}_{-0.04}$ & $294.72^{24.72}_{-25.26}$ & $-16.66$ & $-4.70^{0.01}_{-0.01}$\\
$(23.52,23.77)$& $-1.22^{0.03}_{-0.03}$ & $-6.72^{2.24}_{-2.25}$ & $408.96^{21.98}_{-22.39}$ & $-1.22^{0.03}_{-0.03}$ & $529.61^{41.29}_{-40.89}$ & $-17.99$ & $-4.36^{0.01}_{-0.01}$\\
$(23.77,24.73)$& $-1.27^{0.01}_{-0.01}$ & $-6.93^{2.07}_{-2.10}$ & $522.90^{17.03}_{-16.87}$ & $-1.27^{0.01}_{-0.01}$ & $720.33^{33.10}_{-32.33}$ & $-19.23$ & $-4.36^{0.01}_{-0.01}$\\
$(24.73,24.98)$& $-1.24^{0.02}_{-0.02}$ & $-7.03^{2.06}_{-2.03}$ & $715.33^{38.80}_{-38.17}$ & $-1.24^{0.02}_{-0.02}$ & $942.52^{73.93}_{-72.75}$ & $-18.35$ & $-4.21^{0.01}_{-0.01}$\\
$(24.98,25.03)$& $-1.21^{0.06}_{-0.07}$ & $-6.64^{2.30}_{-2.24}$ & $443.67^{53.11}_{-53.40}$ & $-1.21^{0.06}_{-0.06}$ & $567.77^{97.01}_{-97.82}$ & $-17.13$ & $-4.43^{0.03}_{-0.03}$\\
$(25.03,25.15)$& $-1.21^{0.06}_{-0.06}$ & $-6.86^{2.20}_{-2.21}$ & $291.20^{23.51}_{-23.65}$ & $-1.21^{0.06}_{-0.06}$ & $371.57^{48.53}_{-51.50}$ & $-17.35$ & $-4.67^{0.02}_{-0.02}$\\
$(25.15,25.29)$& $-1.19^{0.04}_{-0.04}$ & $-6.35^{2.32}_{-2.36}$ & $309.51^{18.79}_{-18.62}$ & $-1.19^{0.04}_{-0.04}$ & $387.17^{38.25}_{-37.17}$ & $-17.48$ & $-4.45^{0.01}_{-0.01}$\\
$(25.29,25.65)$& $-1.33^{0.02}_{-0.02}$ & $-7.13^{1.91}_{-1.98}$ & $418.60^{20.94}_{-20.08}$ & $-1.33^{0.02}_{-0.02}$ & $629.77^{44.36}_{-45.50}$ & $-20.63$ & $-4.34^{0.01}_{-0.01}$\\
$(25.65,25.83)$& $-1.39^{0.03}_{-0.04}$ & $-6.92^{2.02}_{-2.03}$ & $364.07^{27.71}_{-28.00}$ & $-1.39^{0.03}_{-0.03}$ & $594.39^{70.40}_{-70.58}$ & $-21.79$ & $-4.48^{0.01}_{-0.01}$\\
$(25.83,25.93)$& $-1.22^{0.04}_{-0.04}$ & $-6.75^{2.32}_{-2.26}$ & $347.23^{23.17}_{-23.19}$ & $-1.22^{0.04}_{-0.04}$ & $449.13^{50.64}_{-48.44}$ & $-17.76$ & $-4.32^{0.02}_{-0.01}$\\
$(25.93,26.11)$& $-1.40^{0.04}_{-0.04}$ & $-6.57^{2.25}_{-2.26}$ & $250.82^{17.61}_{-17.39}$ & $-1.40^{0.04}_{-0.04}$ & $421.96^{51.26}_{-50.74}$ & $-22.13$ & $-4.58^{0.01}_{-0.01}$\\
$(26.11,26.45)$& $-1.37^{0.04}_{-0.04}$ & $-6.68^{2.19}_{-2.24}$ & $140.08^{6.50}_{-6.42}$ & $-1.37^{0.04}_{-0.04}$ & $223.99^{21.54}_{-21.57}$ & $-21.42$ & $-4.82^{0.01}_{-0.01}$\\
$(26.45,26.81)$& $-1.54^{0.03}_{-0.03}$ & $-5.69^{2.47}_{-2.83}$ & $307.46^{22.40}_{-22.36}$ & $-1.54^{0.03}_{-0.03}$ & $677.65^{77.96}_{-77.01}$ & $-28.05$ & $-4.56^{0.02}_{-0.01}$\\
$(26.81,27.88)$& $-1.53^{0.02}_{-0.02}$ & $-7.00^{2.07}_{-2.06}$ & $374.84^{19.07}_{-19.45}$ & $-1.53^{0.02}_{-0.02}$ & $806.50^{63.67}_{-63.96}$ & $-25.77$ & $-4.65^{0.01}_{-0.01}$\\
$(27.88,28.00)$& $-1.44^{0.04}_{-0.04}$ & $-6.88^{2.09}_{-2.15}$ & $410.33^{44.82}_{-44.07}$ & $-1.44^{0.04}_{-0.04}$ & $739.31^{112.31}_{-120.93}$ & $-22.98$ & $-4.52^{0.02}_{-0.02}$\\
$(28.00,28.27)$& $-1.40^{0.02}_{-0.02}$ & $-6.66^{2.24}_{-2.24}$ & $562.40^{36.51}_{-36.02}$ & $-1.40^{0.02}_{-0.02}$ & $940.56^{87.46}_{-87.74}$ & $-22.43$ & $-4.30^{0.01}_{-0.01}$\\
$(28.27,28.44)$& $-1.41^{0.03}_{-0.03}$ & $-6.42^{2.27}_{-2.44}$ & $671.40^{59.39}_{-60.75}$ & $-1.41^{0.03}_{-0.03}$ & $1130.90^{140.13}_{-136.02}$ & $-22.78$ & $-4.32^{0.02}_{-0.02}$\\
$(28.44,28.69)$& $-1.41^{0.03}_{-0.03}$ & $-6.88^{2.14}_{-2.15}$ & $581.09^{51.84}_{-50.83}$ & $-1.41^{0.03}_{-0.03}$ & $996.02^{122.94}_{-120.00}$ & $-22.54$ & $-4.47^{0.02}_{-0.02}$\\
$(28.69,28.99)$& $-1.41^{0.03}_{-0.03}$ & $-5.82^{2.47}_{-2.78}$ & $650.63^{63.42}_{-65.40}$ & $-1.41^{0.03}_{-0.03}$ & $1112.09^{154.89}_{-149.27}$ & $-24.18$ & $-4.55^{0.02}_{-0.02}$\\
$(28.99,29.69)$& $-1.43^{0.02}_{-0.02}$ & $-6.33^{2.35}_{-2.41}$ & $461.56^{32.86}_{-33.00}$ & $-1.43^{0.02}_{-0.02}$ & $805.46^{85.47}_{-85.73}$ & $-23.43$ & $-4.75^{0.01}_{-0.01}$\\
$(29.69,30.14)$& $-1.41^{0.02}_{-0.02}$ & $-7.05^{2.01}_{-2.01}$ & $571.66^{45.09}_{-45.72}$ & $-1.41^{0.02}_{-0.02}$ & $968.02^{105.53}_{-105.79}$ & $-22.38$ & $-4.60^{0.02}_{-0.01}$\\
$(30.14,30.53)$& $-1.37^{0.04}_{-0.04}$ & $-6.80^{2.21}_{-2.21}$ & $423.55^{38.12}_{-37.53}$ & $-1.37^{0.04}_{-0.04}$ & $672.71^{90.49}_{-89.73}$ & $-21.31$ & $-4.79^{0.02}_{-0.02}$\\
$(30.53,31.98)$& $-1.40^{0.03}_{-0.03}$ & $-6.78^{2.16}_{-2.21}$ & $200.28^{8.27}_{-8.44}$ & $-1.40^{0.03}_{-0.03}$ & $333.21^{24.47}_{-24.19}$ & $-22.19$ & $-5.03^{0.01}_{-0.01}$\\
$(31.98,32.65)$& $-1.58^{0.05}_{-0.05}$ & $-6.84^{2.24}_{-2.22}$ & $118.65^{8.49}_{-8.33}$ & $-1.58^{0.05}_{-0.05}$ & $286.51^{40.69}_{-40.73}$ & $-26.88$ & $-5.20^{0.01}_{-0.01}$\\
$(32.65,32.94)$& $-1.64^{0.05}_{-0.06}$ & $-6.10^{2.76}_{-2.71}$ & $232.41^{36.28}_{-39.30}$ & $-1.64^{0.05}_{-0.05}$ & $674.09^{183.06}_{-185.50}$ & $-29.12$ & $-5.00^{0.03}_{-0.02}$\\
$(32.94,33.39)$& $-1.52^{0.03}_{-0.03}$ & $-6.31^{2.43}_{-2.56}$ & $335.24^{30.65}_{-31.57}$ & $-1.52^{0.03}_{-0.03}$ & $706.64^{95.28}_{-96.99}$ & $-25.42$ & $-4.78^{0.02}_{-0.02}$\\
$(33.39,34.02)$& $-1.49^{0.03}_{-0.03}$ & $-6.86^{2.20}_{-2.17}$ & $262.45^{18.67}_{-19.11}$ & $-1.49^{0.03}_{-0.03}$ & $522.17^{63.89}_{-63.48}$ & $-24.60$ & $-4.95^{0.01}_{-0.01}$\\
$(34.02,34.26)$& $-1.42^{0.05}_{-0.05}$ & $-6.74^{2.23}_{-2.25}$ & $207.40^{15.93}_{-16.00}$ & $-1.42^{0.05}_{-0.05}$ & $363.90^{51.90}_{-52.06}$ & $-22.58$ & $-4.87^{0.02}_{-0.02}$\\
$(34.26,34.84)$& $-1.53^{0.03}_{-0.03}$ & $-6.12^{2.50}_{-2.56}$ & $304.32^{26.22}_{-26.69}$ & $-1.53^{0.03}_{-0.03}$ & $655.18^{91.92}_{-92.57}$ & $-26.17$ & $-4.89^{0.02}_{-0.01}$\\
$(34.84,35.10)$& $-1.49^{0.04}_{-0.04}$ & $-6.55^{2.29}_{-2.32}$ & $387.89^{44.24}_{-44.84}$ & $-1.49^{0.04}_{-0.04}$ & $769.11^{128.33}_{-127.88}$ & $-24.38$ & $-4.70^{0.02}_{-0.02}$\\
$(35.10,35.65)$& $-1.48^{0.03}_{-0.03}$ & $-6.20^{2.45}_{-2.56}$ & $361.70^{28.53}_{-28.95}$ & $-1.48^{0.03}_{-0.03}$ & $702.20^{84.41}_{-86.90}$ & $-24.96$ & $-4.80^{0.02}_{-0.01}$\\
$(35.65,35.97)$& $-1.48^{0.06}_{-0.06}$ & $-5.90^{2.73}_{-2.81}$ & $252.30^{32.00}_{-32.27}$ & $-1.49^{0.06}_{-0.06}$ & $509.99^{108.33}_{-112.39}$ & $-25.23$ & $-5.02^{0.03}_{-0.02}$\\
$(35.97,36.23)$& $-1.48^{0.06}_{-0.06}$ & $-6.01^{2.56}_{-2.61}$ & $186.51^{17.37}_{-17.17}$ & $-1.48^{0.06}_{-0.06}$ & $367.26^{61.46}_{-62.98}$ & $-24.46$ & $-4.96^{0.02}_{-0.02}$\\
$(36.23,36.78)$& $-1.43^{0.05}_{-0.05}$ & $-6.55^{2.37}_{-2.32}$ & $141.68^{9.07}_{-8.99}$ & $-1.44^{0.05}_{-0.05}$ & $255.45^{31.37}_{-31.57}$ & $-22.99$ & $-5.11^{0.02}_{-0.01}$\\
$(36.78,37.49)$& $-1.56^{0.03}_{-0.03}$ & $-6.95^{2.00}_{-2.00}$ & $259.93^{19.61}_{-19.10}$ & $-1.56^{0.03}_{-0.03}$ & $597.89^{71.18}_{-73.10}$ & $-26.48$ & $-4.92^{0.01}_{-0.01}$\\
$(37.49,39.04)$& $-1.41^{0.03}_{-0.03}$ & $-6.07^{2.52}_{-2.60}$ & $196.59^{9.06}_{-9.00}$ & $-1.41^{0.03}_{-0.03}$ & $338.34^{28.87}_{-29.01}$ & $-24.16$ & $-5.11^{0.01}_{-0.01}$\\
$(39.04,39.99)$& $-1.52^{0.04}_{-0.04}$ & $-6.69^{2.30}_{-2.23}$ & $200.27^{17.29}_{-17.05}$ & $-1.52^{0.04}_{-0.04}$ & $422.09^{62.90}_{-64.72}$ & $-25.27$ & $-5.28^{0.02}_{-0.02}$\\
$(39.99,40.44)$& $-1.67^{0.05}_{-0.05}$ & $-6.28^{2.50}_{-2.55}$ & $361.82^{77.70}_{-81.98}$ & $-1.66^{0.04}_{-0.04}$ & $995.27^{255.78}_{-242.03}$ & $-29.03$ & $-5.10^{0.03}_{-0.03}$\\
$(40.44,41.65)$& $-1.62^{0.06}_{-0.06}$ & $-6.73^{2.34}_{-2.29}$ & $104.52^{3.79}_{-3.60}$ & $-1.61^{0.06}_{-0.06}$ & $240.56^{44.14}_{-44.55}$ & $-25.84$ & $-5.55^{0.01}_{-0.01}$\\
$(41.65,42.12)$& $-1.80^{0.06}_{-0.06}$ & $-6.15^{2.68}_{-2.64}$ & $137.72^{21.33}_{-24.99}$ & $-1.79^{0.06}_{-0.06}$ & $687.84^{247.22}_{-236.36}$ & $-33.37$ & $-5.25^{0.03}_{-0.03}$\\
$(42.12,42.70)$& $-1.68^{0.07}_{-0.07}$ & $-6.04^{1.97}_{-1.93}$ & $104.74^{3.81}_{-3.88}$ & $-1.67^{0.09}_{-0.09}$ & $241.42^{61.81}_{-62.81}$ & $-22.16$ & $-5.46^{0.02}_{-0.02}$\\
$(42.70,43.22)$& $-1.83^{0.06}_{-0.06}$ & $-5.89^{2.78}_{-2.73}$ & $263.01^{64.26}_{-95.53}$ & $-1.80^{0.05}_{-0.04}$ & $998.23^{296.50}_{-273.90}$ & $-34.68$ & $-5.21^{0.06}_{0.04}$\\
$(43.22,43.73)$& $-1.71^{0.07}_{-0.07}$ & $-6.51^{2.41}_{-2.45}$ & $166.16^{30.49}_{-32.68}$ & $-1.71^{0.07}_{-0.07}$ & $606.40^{210.36}_{-205.64}$ & $-29.76$ & $-5.39^{0.03}_{-0.03}$\\
$(43.73,44.57)$& $-1.64^{0.04}_{-0.04}$ & $-6.61^{2.28}_{-2.31}$ & $201.49^{21.76}_{-22.29}$ & $-1.64^{0.04}_{-0.04}$ & $569.34^{110.68}_{-109.88}$ & $-28.51$ & $-5.23^{0.02}_{-0.02}$\\
$(45.50,45.94)$& $-1.53^{0.09}_{-0.09}$ & $-6.60^{2.32}_{-2.31}$ & $132.85^{16.72}_{-16.91}$ & $-1.54^{0.09}_{-0.09}$ & $299.61^{75.69}_{-75.41}$ & $-25.11$ & $-5.45^{0.03}_{-0.02}$\\
$(45.94,47.99)$& $-1.58^{0.03}_{-0.03}$ & $-6.17^{2.54}_{-2.55}$ & $217.66^{16.67}_{-16.41}$ & $-1.58^{0.03}_{-0.03}$ & $525.28^{63.62}_{-65.32}$ & $-27.60$ & $-5.24^{0.02}_{-0.01}$\\
$(47.99,49.40)$& $-1.52^{0.05}_{-0.05}$ & $-6.58^{2.28}_{-2.36}$ & $150.42^{12.28}_{-12.20}$ & $-1.52^{0.05}_{-0.05}$ & $319.61^{50.97}_{-51.11}$ & $-25.28$ & $-5.45^{0.02}_{-0.01}$\\
$(49.40,49.96)$& $-1.53^{0.06}_{-0.06}$ & $-6.12^{2.67}_{-2.70}$ & $246.13^{36.36}_{-37.36}$ & $-1.53^{0.06}_{-0.06}$ & $551.12^{133.52}_{-138.59}$ & $-25.44$ & $-5.26^{0.03}_{-0.03}$\\
$(49.96,51.15)$& $-1.39^{0.07}_{-0.07}$ & $-5.45^{2.55}_{-2.91}$ & $137.95^{12.57}_{-12.67}$ & $-1.41^{0.06}_{-0.06}$ & $243.04^{39.19}_{-38.73}$ & $-24.36$ & $-5.47^{0.03}_{-0.02}$\\
$(51.15,51.83)$& $-1.60^{0.10}_{-0.10}$ & $-6.60^{2.44}_{-2.39}$ & $105.10^{3.85}_{-4.27}$ & $-1.50^{0.11}_{-0.10}$ & $145.96^{30.36}_{-33.77}$ & $-13.52$ & $-5.67^{0.02}_{-0.02}$\\
$(52.33,53.01)$& $-1.51^{0.09}_{-0.09}$ & $-6.60^{2.33}_{-2.33}$ & $108.19^{6.50}_{-6.51}$ & $-1.47^{0.11}_{-0.11}$ & $184.10^{45.88}_{-47.36}$ & $-22.35$ & $-5.63^{0.02}_{-0.02}$\\
$(53.01,53.77)$& $-1.66^{0.06}_{-0.06}$ & $-6.01^{2.67}_{-2.70}$ & $214.26^{37.29}_{-39.14}$ & $-1.66^{0.06}_{-0.06}$ & $667.97^{199.59}_{-198.95}$ & $-29.31$ & $-5.36^{0.04}_{-0.03}$\\
$(53.77,55.26)$& $-1.40^{0.06}_{-0.06}$ & $-6.33^{2.41}_{-2.46}$ & $114.16^{7.65}_{-7.89}$ & $-1.40^{0.06}_{-0.07}$ & $194.34^{29.38}_{-28.40}$ & $-22.17$ & $-5.57^{0.02}_{-0.02}$\\
$(56.52,59.57)$& $-1.70^{0.05}_{-0.05}$ & $-6.78^{2.13}_{-2.22}$ & $101.85^{1.53}_{-1.53}$ & $-1.63^{0.07}_{-0.07}$ & $176.91^{30.57}_{-31.84}$ & $1.29$ & $-5.74^{0.01}_{-0.01}$\\
$(59.57,61.27)$& $-1.51^{0.06}_{-0.06}$ & $-6.35^{2.47}_{-2.52}$ & $132.65^{11.18}_{-11.25}$ & $-1.51^{0.06}_{-0.06}$ & $279.16^{46.20}_{-47.23}$ & $-24.98$ & $-5.56^{0.02}_{-0.02}$\\
$(61.27,62.73)$& $-1.42^{0.08}_{-0.08}$ & $-6.83^{2.21}_{-2.16}$ & $109.53^{7.01}_{-7.04}$ & $-1.41^{0.08}_{-0.08}$ & $183.40^{32.62}_{-32.96}$ & $-22.08$ & $-5.74^{0.02}_{-0.02}$\\
$(64.17,68.21)$& $-1.63^{0.05}_{-0.05}$ & $-6.44^{2.48}_{-2.47}$ & $135.25^{13.49}_{-13.45}$ & $-1.63^{0.05}_{-0.05}$ & $377.44^{73.17}_{-73.28}$ & $-28.07$ & $-5.79^{0.02}_{-0.02}$\\
$(68.21,72.47)$& $-1.51^{0.06}_{-0.06}$ & $-6.50^{2.39}_{-2.39}$ & $129.03^{12.79}_{-12.58}$ & $-1.51^{0.06}_{-0.06}$ & $269.99^{50.57}_{-51.01}$ & $-24.68$ & $-5.90^{0.02}_{-0.02}$\\
$(79.52,84.33)$& $-1.57^{0.08}_{-0.08}$ & $-6.50^{2.40}_{-2.36}$ & $108.97^{7.16}_{-7.05}$ & $-1.53^{0.10}_{-0.10}$ & $211.32^{55.47}_{-57.77}$ & $-24.28$ & $-6.10^{0.02}_{-0.02}$\\
$(84.33,96.10)$& $-1.72^{0.08}_{-0.08}$ & $-6.51^{2.39}_{-2.44}$ & $107.60^{5.91}_{-6.29}$ & $-1.66^{0.09}_{-0.09}$ & $241.79^{71.07}_{-74.39}$ & $-23.49$ & $-6.30^{0.02}_{-0.02}$\\
\hline  
\end{longtable}

\end{appendix}
\end{document}